# Radial Scaling in Inclusive Jet Production at Hadron Colliders

Frank E. Taylor

Department of Physics
Laboratory of Nuclear Science
Massachusetts Institute of Technology
Cambridge, MA 02139
*fet@mit.edu*

## ABSTRACT

Inclusive jet production in p-p and p̄-p collisions shows many of the same kinematic systematics as observed in single particle inclusive production at much lower energies. In an earlier study (1974) a phenomenology, called *radial scaling*, was developed for the single particle inclusive cross sections that attempted to capture the essential underlying physics of point-like parton scattering and the fragmentation of partons into hadrons suppressed by the kinematic boundary. The phenomenology was successful in emphasizing the underlying systematics of the inclusive particle productions. Here we demonstrate that inclusive jet production at the LHC in high-energy p-p collisions and at the Tevatron in p̄-p inelastic scattering show similar behavior. The ATLAS inclusive jet production plotted as a function of this scaling variable is studied for √s of 2.76, 7 and 13 TeV and is compared to p̄-p inclusive jet production at 1.96 TeV measured at the CDF and D0 at the Tevatron and p-Pb inclusive jet production at the LHC ATLAS at $\sqrt{s_{NN}}$ = 5.02 TeV. Inclusive single particle production at FNAL fixed target and ISR energies are compared to inclusive J/ψ production at the LHC measured in ATLAS, CMS and LHCb. Striking common features of the data are discussed.

*An earlier version of this paper appears as: https://arxiv.org/abs/1704.07341*

*Corresponding email: fet@mit.edu*

# I. INTRODUCTION

Single-particle inclusive productions were studied extensively in the early 1970s as a hadronic analogue to deep inelastic electron-nucleon scattering studies conducted at SLAC. The theoretical underpinnings of single particle inclusive production were developed by Field and Feynman [1], Field, Feynman and Fox [2] and others [3], who described the production of the detected particle to originate from the hard-elastic scattering of a pair of incoming partons which subsequently fragment and hadronize into the inclusively detected particles. The same general quest has been followed in inclusive jet production at hadron colliders (LHC, Tevatron) to test Quantum Chromodynamics (QCD) and to provide the standard model foundation for searches for phenomenon beyond the standard model. In the case of inclusive jet production, incoming partons hard scatter, fragment, then hadronize into cones of particles that form jets where the jet itself is analyzed as the inclusively detected 'particle'.

Some 40 years ago, in the early time of the operation of the Fermilab synchrotron and at the SPS synchrotron and the Intersecting Storage Ring at CERN, single particle inclusive productions, such as $p + p \to \pi^0 + X$, $p + p \to \pi^\pm + X$, $p + p \to K^\pm + X$, were studied [4], [5], [6], [7]. When the data were analyzed in terms of the transverse momentum $p_T$ and the *radial scaling* variable $x_R$, the kinematic form of the Lorentz invariant cross section was greatly simplified. The radial scaling variable is defined by $x_R = E/E_{Max}$, where E is the detected single particle total energy in the center-of-momentum (COM) frame, $E_{Max}$ is the corresponding maximum energy and is roughly $= \sqrt{s}/2$ in the p-p COM frame. The radial scaling variable describes the phase space suppression as the single-particle production approaches the kinematic boundary where $E = E_{Max}$. Note that this suppression is independent of the angle of the emitted particle in the COM frame and depends only on the *radial* distance in energy-momentum space to the kinematic boundary.

The earlier analyses of data indicated that the single particle inclusive cross section $Ed^3\sigma/dp^3$ had power law dependences on the transverse momentum $p_T$ and on the variable $(1-x_R)$ that roughly factorizes in the form:

$$E\frac{d^3\sigma}{dp^3} = F(\sqrt{s}, p_T, x_R) \approx \frac{\alpha}{(\Lambda^2 + p_T^2)^{\frac{np_T}{2}}} (1-x_R)^{n_{xR}} \quad (1)$$

where $\Lambda$ is a transverse mass term, potentially important at low $p_T$, $n_{pT}$ and $n_{xR}$ are the power law indices and $\alpha$ is the parameter that controls the magnitude of the invariant cross section and fixes the dimensions to [momentum]$^{-4}$ [e.g. GeV$^{-4}$]. In principle, all the parameters of Eq.1 can be functions of √s, as well as dependent on the inclusively detected particle. However, in the limited energy range of data analyzed in this earlier work, the s-dependence of the inclusive cross section was found to be mostly in the $x_R$ variable itself; namely, for fixed $p_T$ and $x_R$, the inclusive cross sections were roughly constant as √s was varied. The transverse momentum dependence was found to be approximately independent of the inclusively detected particle, but the (1-$x_R$) dependence varied for different inclusive particle productions. However, more extensive data taken at the ISR showed that there is an overall s-dependence beyond that embodied in the $x_R$ variable [8] and this narrowly-defined radial scaling was violated. Nevertheless, even with this additional s-dependence, the radial scaling formulation was helpful in revealing systematics of the single particle inclusive cross sections.

The question naturally arises whether or not the radial scaling phenomenology has any utility in uncovering systematics in inclusive jet and charm production in p-p and heavy ion collisions at the LHC. After all, the theoretical underpinning of single particle inclusive production and jet inclusive production are the same – namely both are described by hard scattering of incoming partons, followed by fragmentation and hadronization, only in the case of jet production an ensemble of particle carrying the scattered parton momentum are collimated and form a jet. The following questions are therefore quite natural:

- *Is the $p_T$ – dependence of inclusive jet production at the LHC a power law?*
- *How does the $p_T$-dependence of inclusive jet production compare with single particle inclusive production?*
- *Is there a power law dependence in (1-$x_R$) as was observed in single particle inclusive processes such as the one given in Eq. 1?*

On general grounds, one may expect that all the parameters of this simple formulation (α and the power law indices $n_{pT}$ and $n_{xR}$) will depend on √s and that there would be no simplification in the much more complex process of inclusive jet production at TeV energies. Nevertheless, it is interesting to seek answers to these questions.

## II. JET COLLIDER DATA

There is now agreement between pQCD calculations to NLO and inclusive jet production at the LHC to better than ~ 20%, except at high rapidity and high $p_T$, in effect 'explaining' the jet production in terms of scattered partons and subsequent scattered parton hadronizations [9]. It is a success of the underlying theory that the simulations based on pQCD calculations show such good agreement. Further improvements in the data-theory agreement are expected with the future consideration of higher order effects [10] and better methods of calculation various sub-processes such as by amplitude methods [11].

Inclusive jet production at hadron colliders is conventionally described by $p_T$ and the rapidity $y = \frac{1}{2}\ln\left(\frac{E+p_z}{E-p_z}\right)$ which is roughly equal to the pseudo-rapidity defined by $\eta$ = - ln [tan (θ/2)], where θ is the angle of the emitted jet in the p-p COM frame with respect to the incoming beams. The invariant inclusive jet (single particle) cross section can by written in terms of the jet transverse momentum, $p_T$, and jet rapidity, y, after integrating over the azimuthal angle ϕ as:

$$E\frac{d^3\sigma}{dp^3} \to \frac{d^2\sigma}{p_T dp_T dy} \quad (2)$$

In this formulation, the invariant cross section is a function of three variables, the COM energy √s, the transverse momentum $p_T$[1] and the rapidity y. The jet mass has been integrated into the rapidity variable y through the value of the jet total energy E.

The invariant cross section for inclusive jet or single particle production could just as well be written in terms of other groupings of three variables such as √s, $p_T$ and a combination of y, $p_T$ and √s assembled together to express the radial scaling variable $x_R$. In the limit of high energy and small particle or jet mass with respect to √s, the radial scaling

---

[1] We have chosen the $p_T$ differential to be $p_T dp_T$ rather than $2p_T dp_T = dp_T^2$

variable $x_R \approx 2\, p_T \cosh(\eta)/\sqrt{s}$, where $\eta$ is the pseudo-rapidity of the jet in the COM frame. Note that $\cosh(y) \sim \cosh(\eta) = 1/\sin(\theta)$ – hence $x_R \sim 2\, p/\sqrt{s} \sim E^*/E^*_{max}$. Fig. 1 shows the relations of the radial scaling variable $x_R$ to $p_T$ to $\eta$ for $\sqrt{s}$ = 13 TeV. Note that lines of constant $\eta$ ($\eta \sim y$) mix $p_T$ and the scaling variable $x_R$. Thus, the kinematic boundary suppression, controlled by $x_R$ is convoluted with the $p_T$ and $\eta$ (y) dependence.

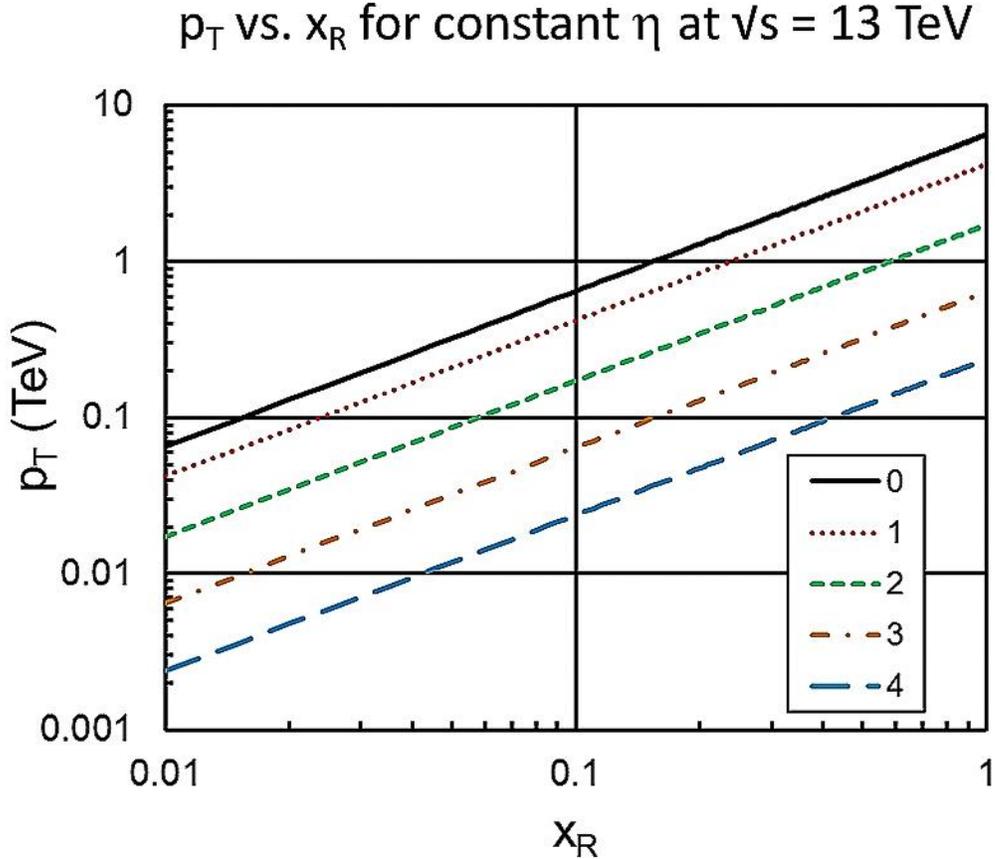

Figure 1: The lines of constant $\eta$ are plotted on the $p_T$-$x_R$ plane for $\sqrt{s}$ = 13 TeV starting at top $|\eta|$ = 0, 1, 2, 3, 4. The region above and to the left of the $\eta = 0$ line (solid black) is kinematically forbidden. Holding $\eta$ (y) constant mixes $p_T$ and $x_R$ and therefore does not control the radial distance from the kinematic limit $x_R = 1$.

## A. Inclusive jets at the LHC

It is interesting to analyze inclusive jet production at the LHC in the simplest terms by seeing if there are kinematic generalities like those observed in the single particle production in p-p collisions. As a typical example, Fig. 2 shows the inclusive jet production at 13 TeV measured by the ATLAS collaboration at the LHC [12], [13], [14].

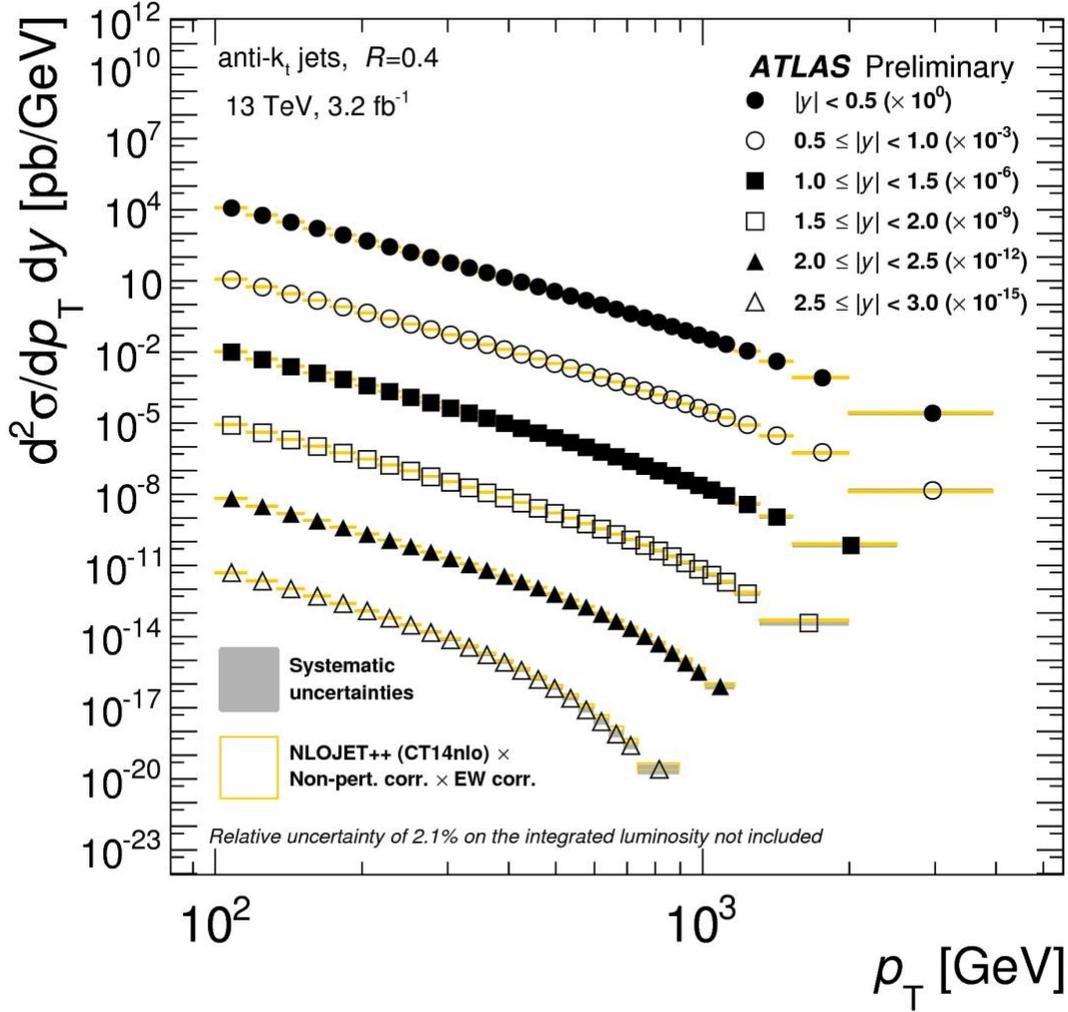

Figure 2: The ATLAS inclusive jet cross section [12] is plotted as a function of $p_T$ for various rapidity regions. The data were taken at $\sqrt{s}$ = 13 TeV and the jets defined by the anti-$k_T$ algorithm. Note that the suppression when close to the kinematic boundary is evident on the RHS of the plot, e.g. for the highest rapidity bin 2.5< |y| <3.0 the maximum $p_T$ is less than 1 TeV/c. The data are in good agreement with simulations.

It is evident that the inclusive jet cross sections agree with the *NLOJET++(CT14nlo)* [15], [16] and corrections. However, we note that plotting the cross section for <u>constant y</u> as a function of $p_T$, as in Fig. 2, involves <u>changing the value of $x_R$</u> as was demonstrated in Fig. 1. Therefore, the presentation of the data for constant y obscures a putative power law behavior in $p_T$ and (1-$x_R$) that we would expect if inclusive jet production in p-p scattering has a similar behavior to that of single particle inclusive cross sections.

Examining Fig. 2, it is evident that the cross section decreases with increasing $p_T$ and y, but it is not obvious that there are any power laws in $p_T$ and $(1-x_R)$ as were observed in single particle inclusive production in p-p collisions. However, we can roughly test the hypothesis that the invariant cross section has the factorized form of Eq. 1 by plotting the resultant invariant cross section $d^2\sigma/p_T dp_T dy$ multiplied by a function of $p_T$, where we find $\sim p_T^6$ works reasonably well. The resultant behavior is shown in Fig. 3. Notice that $p_T^6(d^2\sigma/p_T dp_T dy)$ is mostly a function of $x_R$ in the sense that the data for different values of |y| fall on top of each other and therefore tend to *radially scale for a fixed $\sqrt{s}$*.

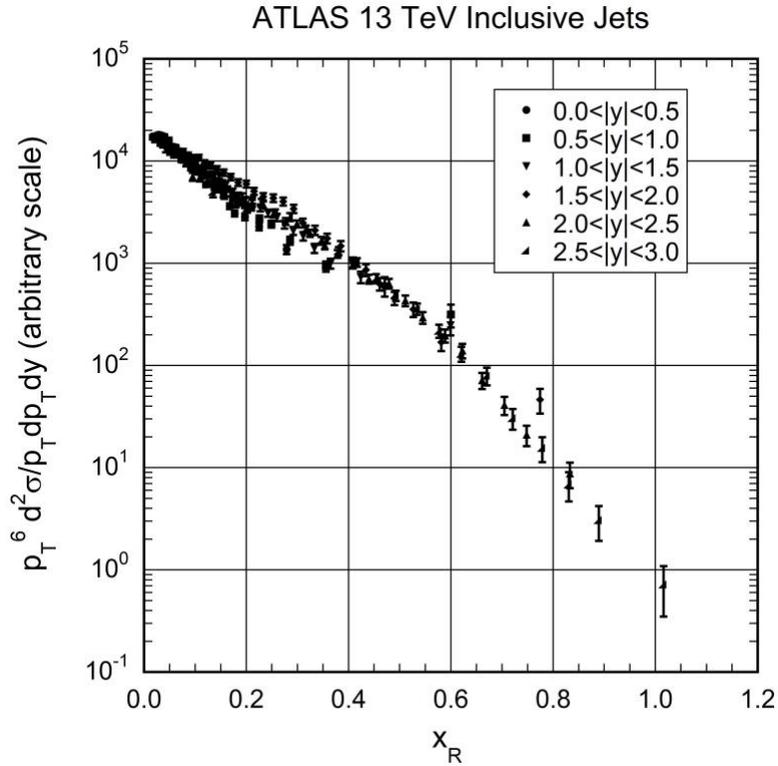

Figure 3: The 13 TeV ATLAS inclusive jet invariant cross section multiplied by $p_T^6$ is plotted as a function $x_R$ for various |y| values. Note that the data tend to roughly scale as a function of $x_R$. The error bars represent the statistical and systematic errors added in quadrature. The point with $x_R > 1$ is due to the finite bin corrections in $p_T$ and y not performed. The data so plotted roughly follow $(1-x_R)^{4.5}$.

For a deeper view of the $p_T$ and $x_R$ dependencies of the 13 TeV inclusive jet data, we analyze the cross section using the form suggested by single-particle inclusive data:

$$\frac{d^2\sigma}{p_T dp_T dy} = A(p_T, s)(1-x_R)^{nx_R} \tag{3}$$

by plotting the cross section in slices of constant $p_T$ as a function of $1-x_R$. In Eq. 3 we have not assumed a specific form for the $p_T$ function, $A(p_T,s)$, except to posit that it is <u>not</u> a function of $x_R$. Since the 13 TeV ATLAS inclusive jet data are binned in rapidity (y), thereby requiring the jet mass to be known in order to determine the angle of the jet in the p-p COM, we approximate the radial scaling variable as:

$$x_R = \frac{E}{E_{max}} = \frac{2\sqrt{\left(p_T^2 \cosh^2(y)(1+(m_J^2/p_T^2)\tanh^2(y)) + m_J^2\right)}}{\sqrt{s}}$$
$$\approx \frac{2p_T \cosh(y)}{\sqrt{s}} \sqrt{\left(1 + \frac{m_J^2}{p_T^2}\tanh^2(y)\right)} \tag{4}$$

where the jet mass $m_J$ has been adsorbed in the variable y but is bounded using the prescription of reference [17] by $m_J/p_T < R/\sqrt{2} = 0.28$ for the jet cone size $R = 0.4$ given by:

$$R = \sqrt{(\Delta\phi^2 + \Delta\eta^2)} \tag{5}$$

where $\Delta\phi$ and $\Delta\eta$ are the jet cone widths in $\phi$ and $\eta$, respectively, defined with respect to the colliding beams axis. The finite y bin size was treated by assuming that the published data value for a bin correspond to the midpoint of the lower and upper limits − a valid assumption for low y, where the rapidity distribution is approximately flat. However, the bin center so calculated will be slightly larger by < 1.4% from a more valid data-weighted value for the highest rapidity bin (2.5 < |y| < 3.0) resulting in the computed value of $x_R$ larger by < 3.8%. This putative finite bin correction was ignored.

The 13 TeV inclusive jet data so analyzed are shown in Fig. 4. We would expect that the $x_R$ behavior would be complicated and, even if a power law were operative, the indices $n_{pT}$ and $n_{xR}$ would be functions of $\sqrt{s}$, $p_T$ and y. Note that $x_R = 0$, where $A(p_T,s)$ is evaluated in the $(1-x_R)$ power law fits corresponds to the limit when $\sqrt{s} \to \infty$ for constant $p_T$ and thus is beyond the minimum value ($\eta = 0$) of $x_{Rmin} = 2p_T/\sqrt{s}$ for finite $\sqrt{s}$. This small extrapolation assumes that the functional form of Eq.3 is valid in the small region from $x_{Rmin}$ to $x_R = 0$.

The power law fits of $(1-x_R)^{n_{xR}}$ were performed by a least-squares linear method on the natural logarithms of the cross section as a function of the $\ln(1-x_R)$ using the statistical and systematic errors added in quadrature. The slopes of these linear fits are the exponents $n_{xR}$ and the constant terms are the logs of $A(p_T,s)$ for the fixed $p_T$ values. In general, we would expect that $n_{xR}$ would be a function of $p_T$ and $\sqrt{s}$.

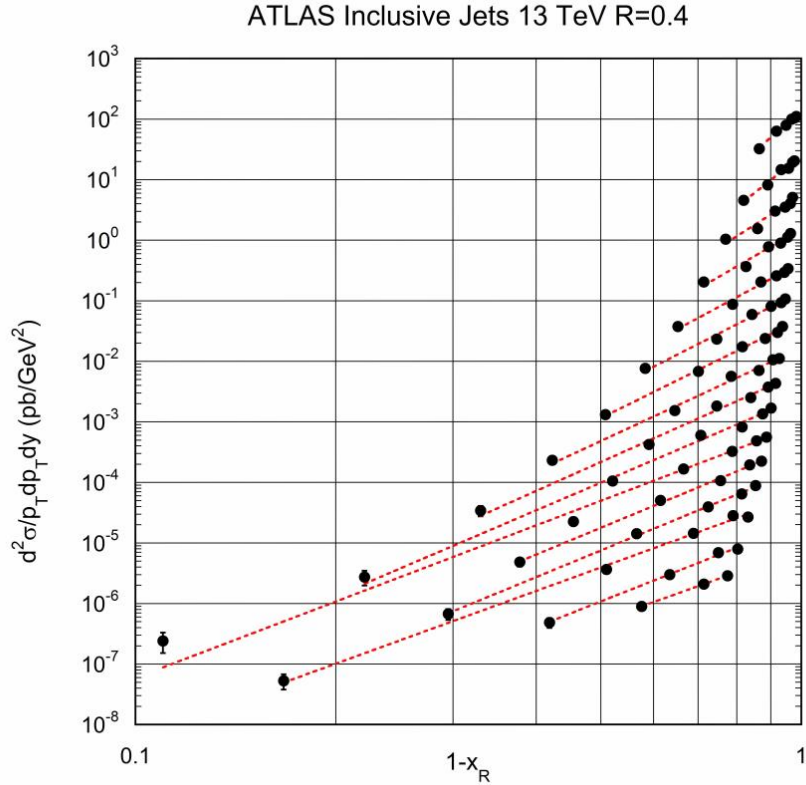

Figure 4: The 13 TeV ATLAS inclusive jet cross section is plotted as a function of $1-x_R$ for various constant values of $p_T$. For clarity, only every second value of constant $p_T$ of the data set is plotted. Starting at the top of figure the lines of constant $p_T$ are: $p_T$ = 0.11, 0.14, 0.18, 0.23, 0.28, 0.33, 0.39, 0.46, 0.53, 0.62. 0.71, 0.81, 0.92, 1.1, 1.2, 1.4 TeV from the top of the figure, respectively. The dotted red lines are power-law fits of the form $A(p_T,s)$ $(1-x_R)^{n_{xR}}$ described in the text. Note that the data are consistent with a power law in $(1-x_R)$ as in the case of single particle inclusive cross sections measured at much lower energies but that the power law indices nxR are a function of $p_T$ - the $(1-x_R)$ power index is larger for lower $p_T$. The displayed error bars are statistical and systematic added in quadrature. The overall error in the luminosity normalization has been neglected.

Fig. 5 shows the values of $n_{xR}$ plotted as a function of $p_T$ where it is evident that $n_{xR} \to \sim 4$ for high $p_T$ but has a higher value for low $p_T$. Fig. 6 is a plot of $A(p_T,s)$ as a function of $p_T$ where it is clear that the data follow a power law as suggested by early radial scaling studies of single particle inclusive scattering denoted by Eq. 1 above. The fits, represented by the dotted red lines in the figures below, have the following forms:

$$n_{xR}(p_T,s) = n_{xR0} + \frac{D(s)}{p_T} \qquad (6)$$

$$A(p_T,s) = \frac{\alpha(s)}{p_T^{n_{pT}}}, \qquad (7)$$

where $n_{xR0}$ and $D$ are the fit parameters for the power of $(1-x_R)$; and $\alpha$ and $n_{pT}$ are the fit parameters for the power law fit to $A(p_T,s)$. Note that at the high $p_T$ values of these data the $\Lambda$ term was not necessary.

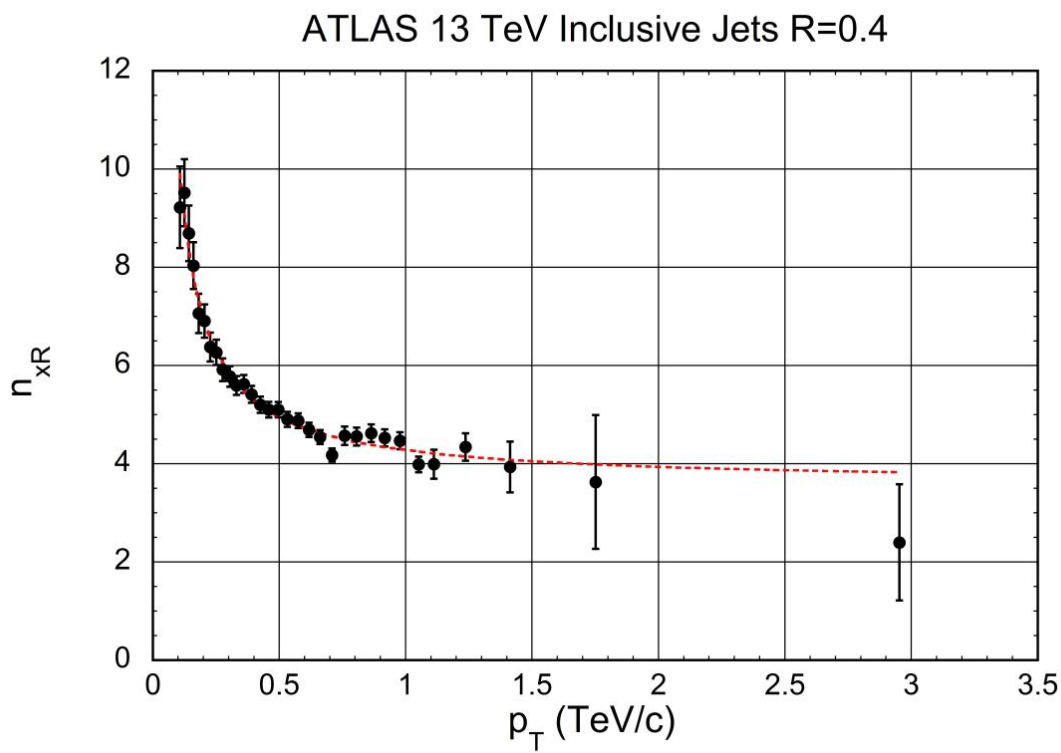

Figure 5: The exponent of the $(1-x_R)$ power law is shown as a function of $p_T$. The red dotted line indicates the fit described in the text that is given by Eq. 6. A $1/p_T$-dependence plus a constant term is evident.

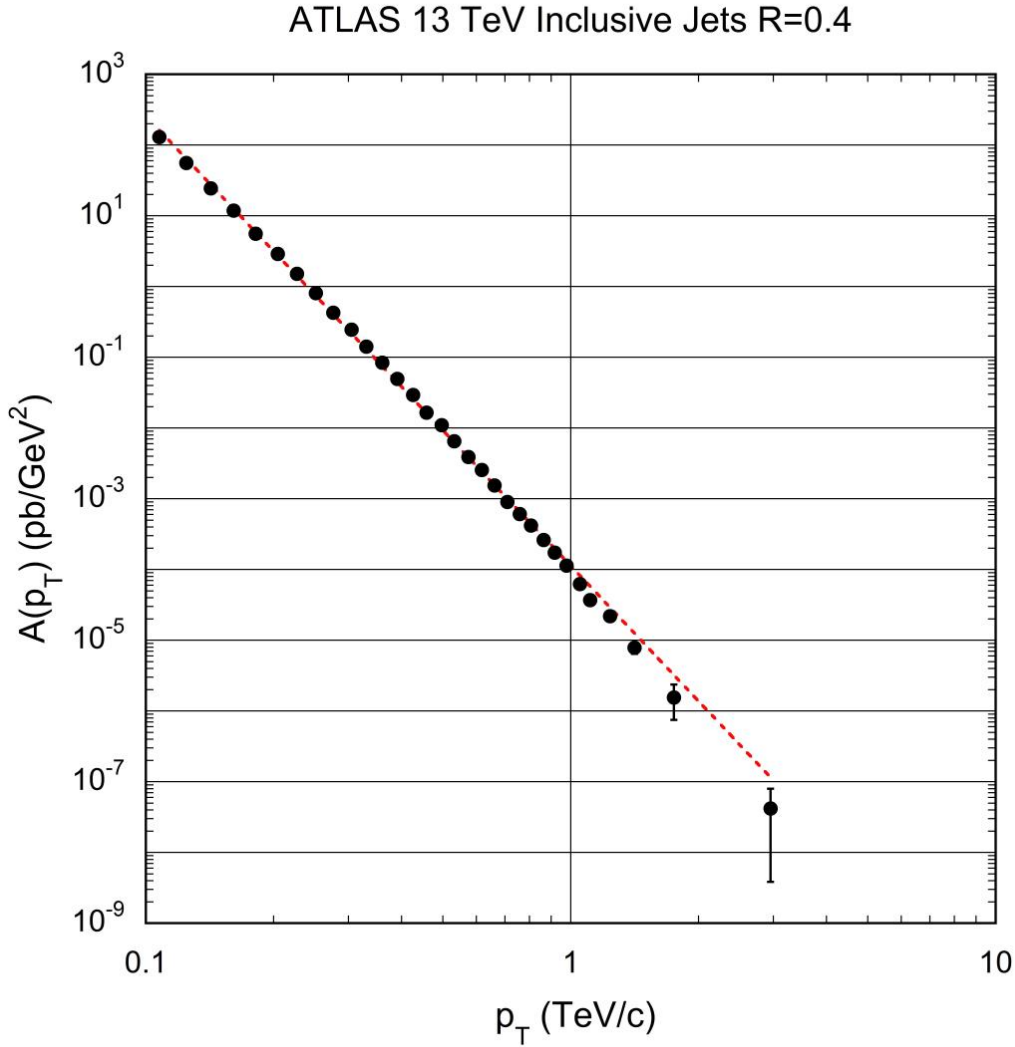

Figure 6: The fit values $A(p_T,s)$ are plotted as a function of $p_T$. Note the power law behavior over nine orders of magnitude. The red dotted line indicates the fit $A(p_T,s) \sim 1/p_T^{6.4}$. Note that the $p_T$-power index (6.4) will be independent of the experimental jet energy scale calibration so long as the energy scale does not depend on the jet energy itself. A small ($\leq \pm 30\%$) deviation from the power law is visible and will be discussed later.

Encouraged by the simplicity of the 13 TeV inclusive jet cross section when analyzed with the radial scaling variable, we now examine the ATLAS jet data taken at $\sqrt{s}$=2.76 [18] and 7 TeV [19]. Both analyses at these lower energies used the same anti-kT jet defining algorithm as well as the same jet cone definition of R = 0.4. The resulting power-law plots are shown in Fig. 7 for the (1-$x_R$) exponent behavior, where we have plotted the exponents as a function of $1/p_T$ to emphasize the linear behavior in that variable, and in Fig. 8 for the $A(p_T,s)$ function. For comparison, the 13 TeV data are plotted on the same scale. Note that the $1/p_T$ term of the $n_{xR}$ dependence grows with increasing s, but the $p_T$ power law exponent, $n_{pT}$, is constant. The overall magnitude of the cross section, governed by the α-term, increases with s.

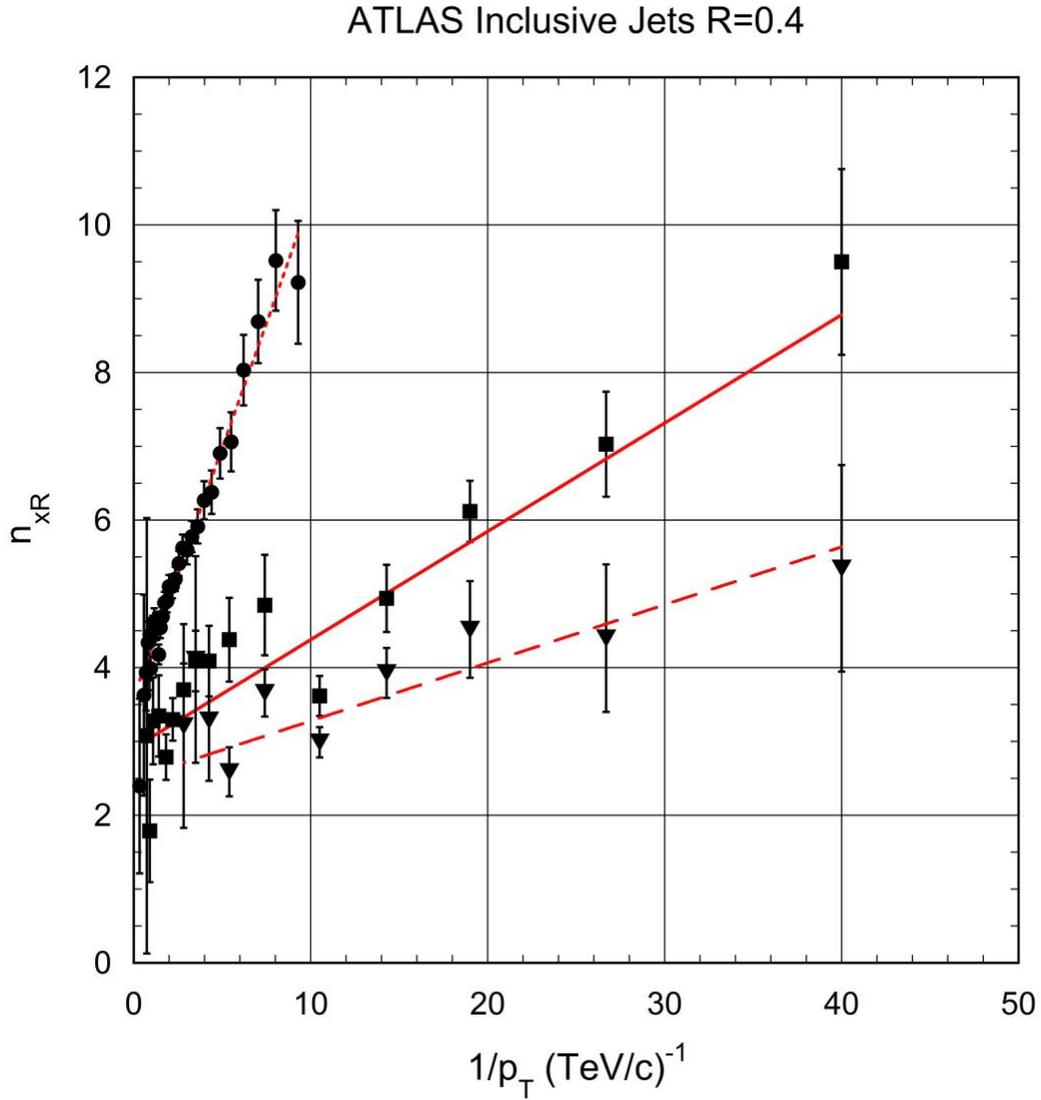

Figure 7: The exponents of the $(1-x_R)$ power law fits are plotted as a function of $1/p_T$. The ATLAS inclusive jets at 13 TeV are represented by circles, 7 TeV by squares and 2.76 TeV by triangles. The red lines are the straight-line fits in $1/p_T$ of the form given by Eq. 6. Only points with $x_R < 0.9$ were considered. The power indices are functions of $p_T$ and $\sqrt{s}$ growing with increasing s.

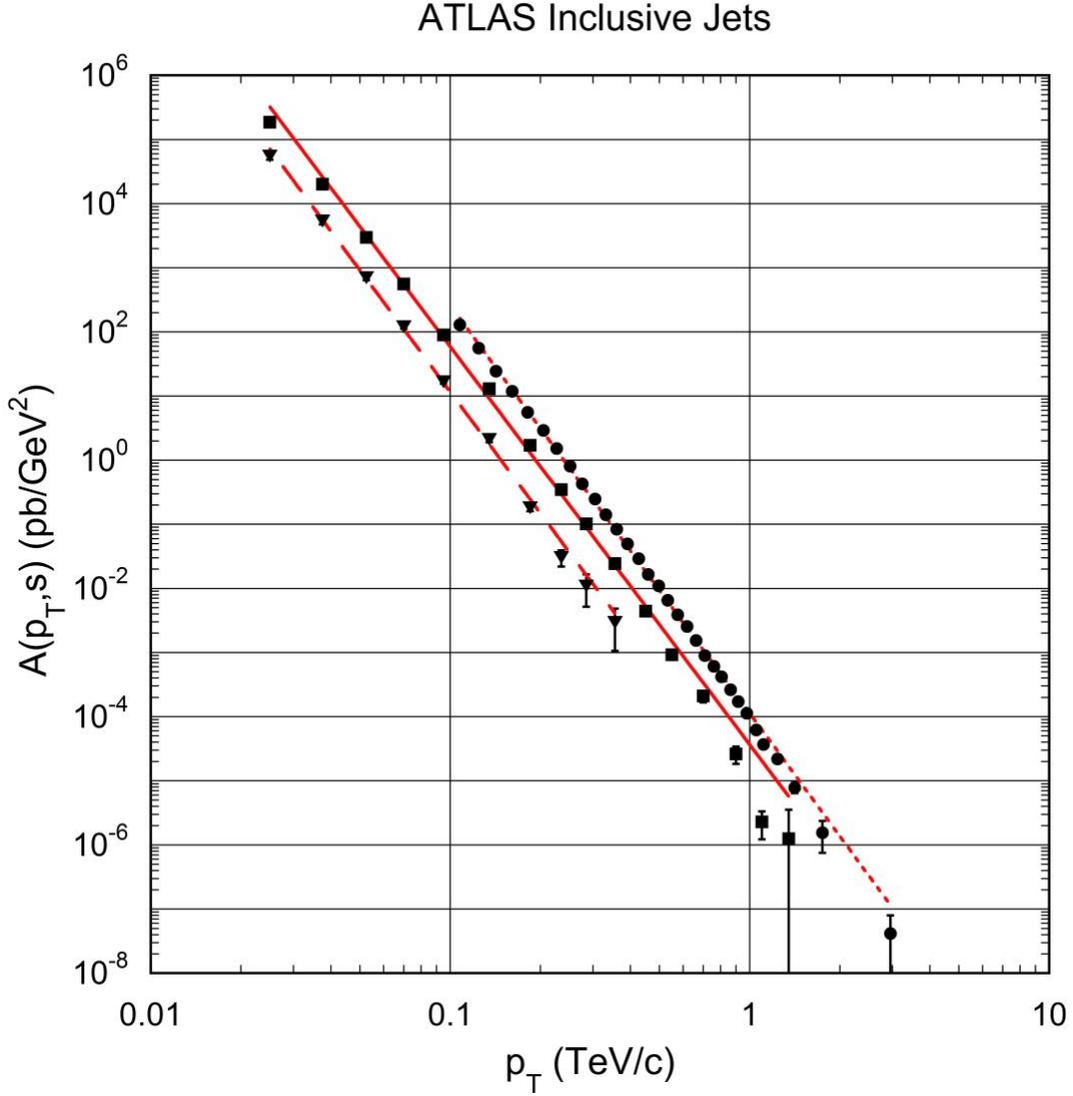

Figure 8: The p$_T$ power law of ATLAS inclusive jet production at the LHC. The jets at 13 TeV are represented by circles, 7 TeV by squares and 2.76 TeV by triangles. The red lines are the power law fits of the form of Eq. 7. We observe that A(p$_T$,s) functions for the three energies have the same power index, but the overall magnitude of A(p$_T$,s) grows with increasing s.

## B. CDF and D0 inclusive jet $\bar{p}$ p data

The CDF [20] and D0 [21] inclusive jet data taken at 1.96 TeV collisions $\bar{p}$-p were analyzed in the same way as the ATLAS inclusive jet data. The results are shown in Figs. 9 and 10.

The power indices of (1-$x_R$) tend to be flatter in rough agreement with the trend seen in Fig. 7, that is, the 1/$p_T$ slope of $n_{xR}$ assumes a smaller value for lower COM energies. In Fig. 10 we notice the same $p_T$ power law behavior as seen in p-p inclusive jets. Hence, we conclude that the p-$\bar{p}$ jets have a behavior consistent with trends shown in Figs. 7 and 8 for the LHC p-p jets.

The fit parameters of the data using Eqs. 6 and 7 are shown in Figs 7 → 10 are given in Tables Ia and Ib below where we have added a measurement by the CMS collaboration of inclusive jets at 8 TeV [22] and 13 TeV [23] to the ATLAS measurements discussed above. Notice that the quality of the $n_{xR}$ fit is reasonable ($\chi^2 < 1.7$/d.f.), whereas that of the power law fit to A($p_T$,s) has a large $\chi^2$/d.f. This will be discussed later.

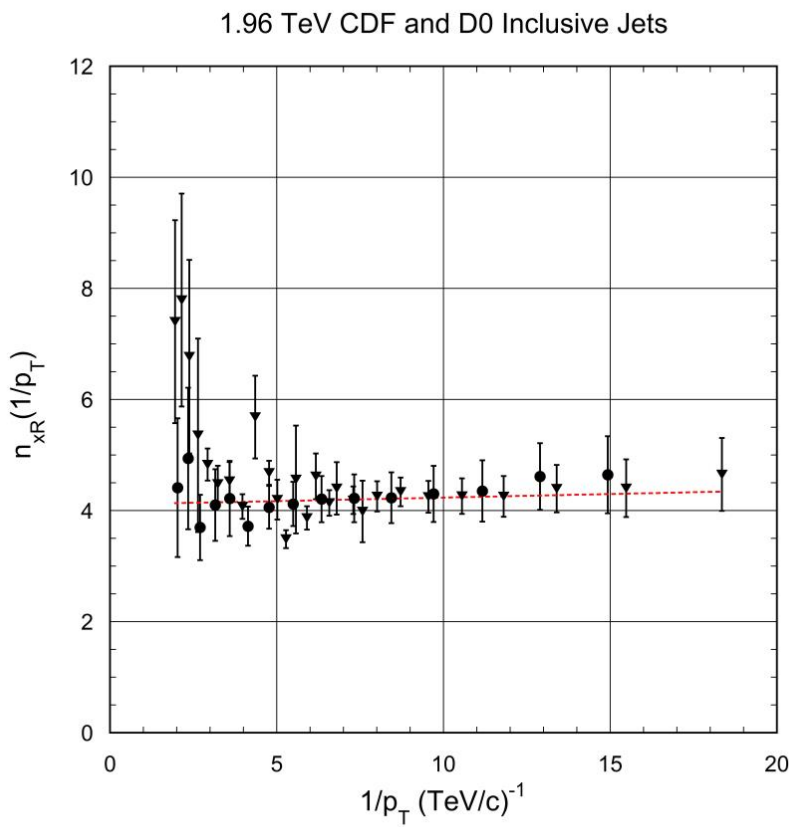

Figure 9: The exponents of the (1-$x_R$) power law fits are plotted as a function of 1/$p_T$ for the CDF (circles) and D0 (triangles) inclusive jet production at 1.96 TeV in $\bar{p}$-p collisions. The dotted red line is the straight-line fits in 1/$p_T$ of the form given by Eq. 6 to the CDF and D0 data considered as one data set. The data show considerable scatter, especially at high $p_T$ (low 1/$p_T$). Only data for $x_R < 0.9$ were included in the fits.

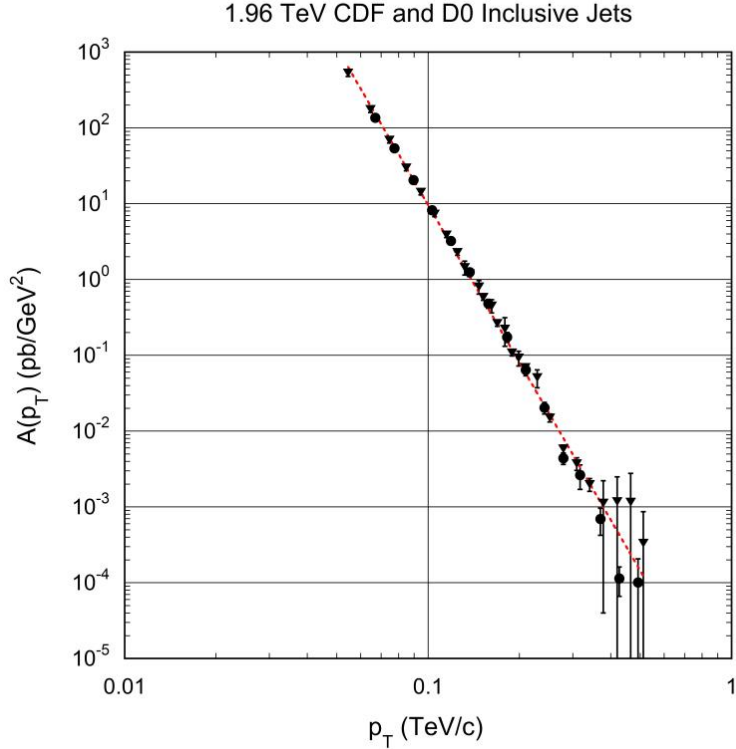

Figure 10: The $p_T$ power law of CDF (circles) and D0 (triangles) for inclusive jet production at the FNAL at 1.96 TeV $\bar{p}$-p collisions. The systematic and statistical errors have been added in quadrature. The dotted red line is a power law fit to both data sets taken together. Only data with $x_R < 0.9$ were considered. The parameters of the fit are in Table Ib.

$N_{XR}$ FITS PP AND P$\bar{P}$: TABLE Ia

| √s (TeV) | D (TeV$^{-1}$) | $n_{xR0}$ | $\chi^2$/d.f. | d.f. |
|---|---|---|---|---|
| 1.96 p̄-p CDF | 0.06 ± 0.04 | 3.7 ± 0.3 | 0.2 | 13 |
| 1.96 p̄-p D0 | 0.00 ± 0.02 | 4.2 ± 0.2 | 2.0 | 25 |
| 2.76 p-p ATLAS | 0.08 ± 0.03 | 2.5 ± 0.3 | 1.2 | 8 |
| 5.02 p-Pb p-side ATLAS | 0.07 ± 0.02 | 3.2 ± 0.2 | 0.8 | 13 |
| 7 p-p ATLAS | 0.15 ± 0.02 | 2.9 ± 0.2 | 1.7 | 14 |
| 8 p-p CMS | 0.22 ± 0.01 | 2.96 ± 0.03 | 1.2 | 33 |
| 13 p-p ATLAS | 0.68 ± 0.03 | 3.61 ± 0.07 | 0.8 | 30 |
| 13 p-p CMS | 0.34 ± 0.09 | 3.5 ± 0.2 | 0.3 | 27 |

Table Ia: The parameters of the fits of the form of Eq. 6 of the power law indices of the $(1-x_R)^{nxR}$ are tabulated.

Examining Tables Ia,b we conclude that most of the variation of the parameters of these fits to inclusive jet production are in the overall normalization term controlled by the parameter α, which increases with increasing √s, and the $p_T$ dependence of the power of (1-$x_R$) given by the term, D, which also increases with increasing √s. The parameters $n_{pT}$ and $nx_R0$ do not show such large systematic √s - dependences.

$P_T$ FITS PP AND P$\bar{P}$: TABLE Ib

| √s (TeV) | α (pb/GeV$^2$) TeV$^{npT}$ | $n_{pT}$ | $\chi^2$/d.f. | d.f. |
|---|---|---|---|---|
| 1.96 p̄-p CDF | (0.9 ± 0.2) x 10$^{-6}$ | 7.03 ± 0.08 | 4 | 13 |
| 1.96 p̄-p D0 | (1.3 ± 0.1) x 10$^{-6}$ | 6.90 ± 0.05 | 1.2 | 25 |
| 2.76 p-p ATLAS | (6.0 ± 1.0) x 10$^{-6}$ | 6.29 ± 0.06 | 3.4 | 8 |
| 7 p-p ATLAS | (3.7 ± 0.2) x 10$^{-5}$ | 6.21 ± 0.03 | 32 | 14 |
| 8 p-p CMS | (2.98 ± 0.04) x 10$^{-5}$ | 6.73 ± 0.01 | 28 | 33 |
| 13 p-p ATLAS | (1.13 ± 0.02) x10$^{-4}$ | 6.36 ± 0.01 | 8 | 30 |
| 13 p-p CMS | (1.06 ± 0.04) x10$^{-4}$ | 6.40 ± 0.03 | 2 | 27 |

Table Ib: The parameters of the power law fits to A($p_T$,s) according to Eq. 7 are tabulated. The $\chi^2$/d.f. values are not very likely and will be discussed later.

The parameters shown in Tables Ia,b are plotted in Fig. 11 below. It is interesting to note that the D and α terms grow linearly with s, whereas the $nx_R0$ and $n_{pT}$ terms are roughly constant. The units of α in Table Ib are [pb/GeV$^2$ TeV$^{npT}$], which for $n_{pT}$ ~ 6 become [energy$^2$] – the same units as the Mandelstam variable, s. Hence it is not surprising that α grows linearly with increasing s in order to preserve the dimensions of the invariant cross section $d^2\sigma/p_T dp_T dy$ to be [pb/GeV$^2$].

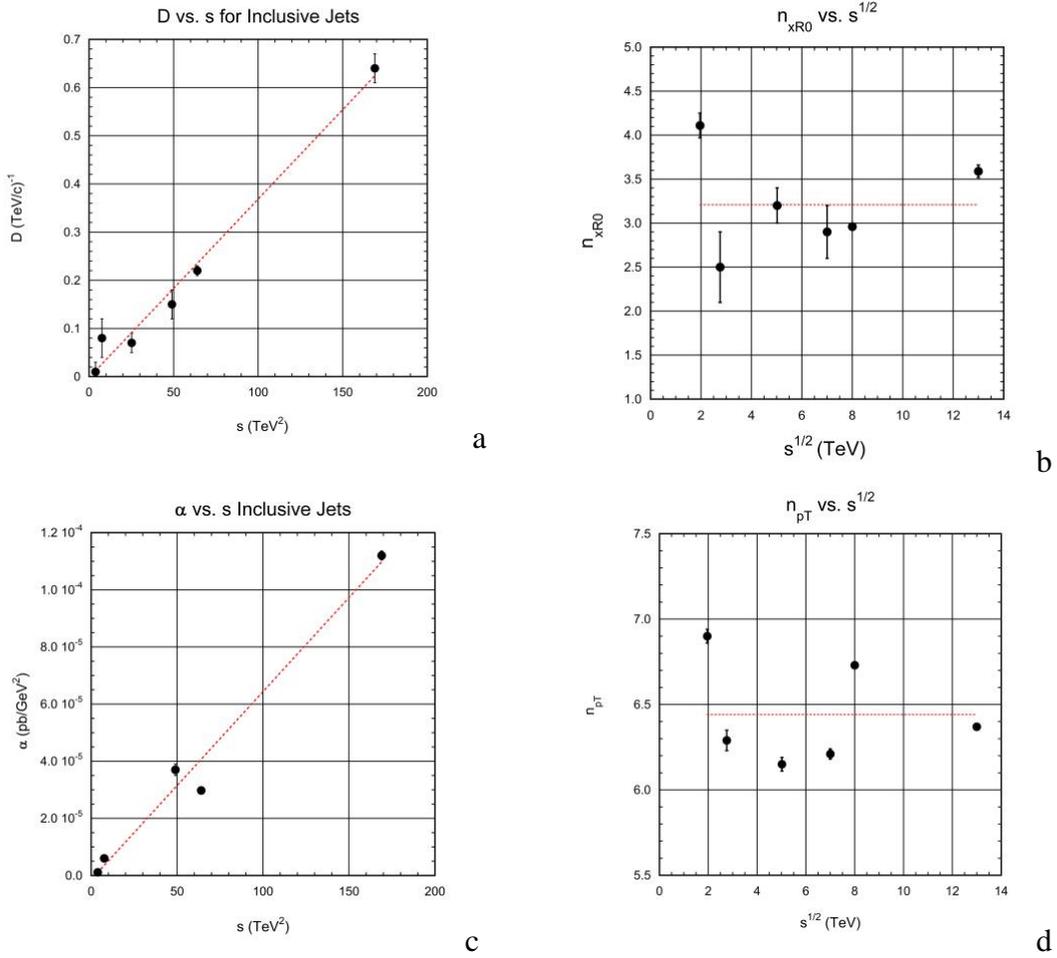

Figure 11: (a,b) The 1-$x_R$ power index parameters are plotted vs. √s. (a): D(s) appears to grow linearly with s as indicated by the red-dotted line; whereas (b) $n_{xR0}$ is roughly constant (red dotted line indicates the average); (c) The cross section magnitude α(s) is plotted, which scales linearly in s, (d) $n_{pT}$ is constant where the red dotted line indicates the average. The CDF and D0 parameters at 1.96 TeV have been combined by weighted average by their respective errors. The 13 TeV ATLAS and CMS values are combined similarly. The p-Pb values at √s = 5.02 TeV [24] have are shown except in the α(s) plot, where no value can be determined since the cross section was self-normalized.

## C. Inclusive jet production in p-Pb collisions

For another view of inclusive jet production at the LHC we analyze the jet data taken in p-Pb collisions at a nucleon-nucleon COM energy of $\sqrt{s_{NN}}$ = 5.02 TeV [24]. Here we examine the two sides of the collision separately – namely the side where the incoming proton fragments (y > 0) and the side where the Pb nucleus fragments (y < 0). Rather than arbitrarily assigning the central rapidity bin -0.3 < y < 0.3 to either the proton forward or the Pb forward data, this central bin was included in both sides of the data. If there were an underlying hard parton-parton scattering that initiates the formation of the detected jet, we would naively expect the same power law in the transverse momentum $p_T$ as observed in jet production in p-p collisions. On the other hand, the fragmentation part of the cross section expressed by the $x_R$-dependence, may be different since the jet formation on the Pb fragmentation side would have to contend with many nucleus fragments, whereas the jet formation on the proton side would be similar to p-p scattering. A difference would be an expression of the well-established jet quenching [25] observed in heavy ion collisions.

The comparison of inclusive jets in p-Pb scattering of the power of $(1-x_R)$ for the two fragmentation cases is shown in Fig. 12. The corresponding $p_T$ dependences are shown in Fig. 13. The fit parameters are listed in Tables IIa and IIb.

### $N_{XR}$ FITS P-P$_B$: TABLE IIa

| √s (TeV) | D (TeV$^{-1}$) | $n_{xR0}$ | $\chi^2$/d.f. | d.f. |
|---|---|---|---|---|
| 5.02 p-side | 0.07 ± 0.02 | 3.2 ± 0.2 | 0.8 | 13 |
| 5.02 Pb-side | 0.3 ± 0.1 | 2.9 ± 0.5 | 0.6 | 10 |

Table IIa: The parameters of the fits of the form of Eq. 6 of the power law indices of the $(1-x_R)^{nxR}$ of constant $p_T$ of Eq. 6 are tabulated. Notice that the $p_T$ dependence in the D-term for Pb forward data is four times larger (roughly three standard deviations) than that of the p-forward case, whereas the $n_{xR0}$ value is the same within errors.

### $P_T$ FITS P-P$_B$: TABLE IIb

| √s (TeV) | $n_{pT}$ | $\chi^2$/d.f. | d.f. |
|---|---|---|---|
| 5.02 p-side | 6.15 ± 0.04 | 25 | 13 |
| 5.02 Pb-side | 6.43 ± 0.07 | 6 | 10 |

Table IIb: The parameters of the power law fits to $A(p_T,s)$ according to Eq. 7 are tabulated. The power indices $np_T$ are the same within 5% for the p-forward and Pb-forward cases and are consistent with the index for p-p scattering given in Table Ib.

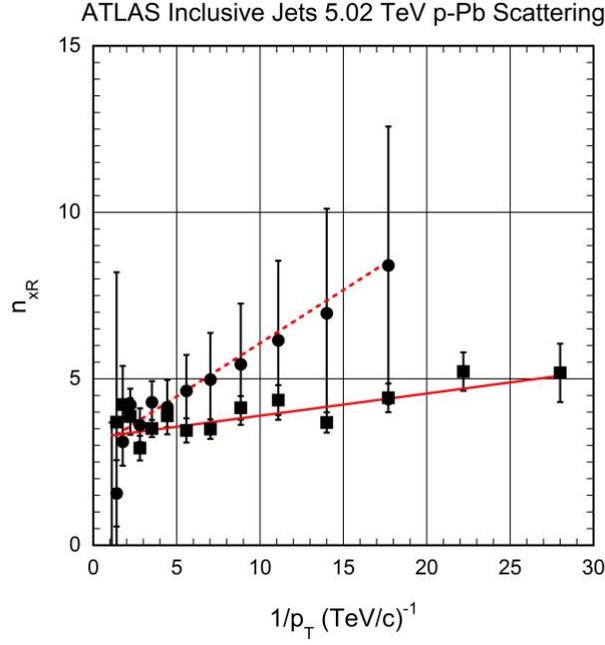

Figure 12: The exponents of the (1-$x_R$) power law fits for the two sides of the ATLAS p-Pb inclusive jet data taken at $\sqrt{s_{NN}}$ = 5.02 TeV. The closed circles correspond to the Pb-forward data and the closed squares to the p-forward side. The red lines are the fits of the form given by Eq. 6. Note that the Pb fragmentation side has a steeper $1/p_T$ dependence than the proton fragmentation side. The error bars represent the statistical and systematic errors added in quadrature. The systematic errors dominate.

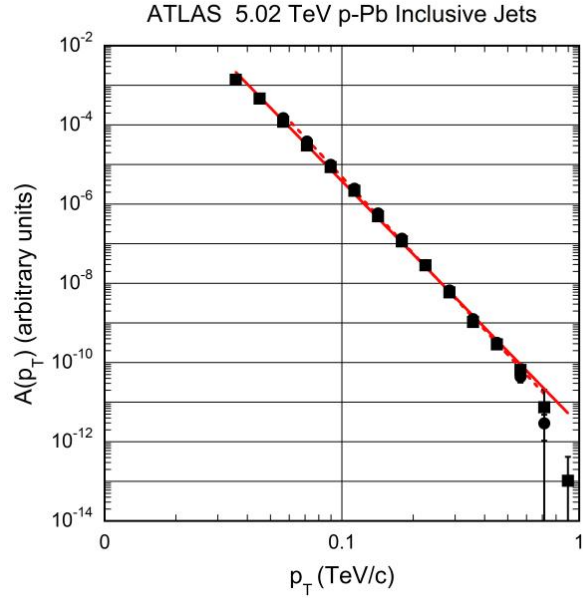

Figure 13: The $p_T$ power law and fits for the two sides of the ATLAS p-Pb inclusive jet data taken at $\sqrt{s_{NN}}$ = 5.02 TeV. The closed circles correspond to the Pb-forward data and the closed squares to the p-forward side. The red lines are the fits of the form given by Eq. 7. Note that the $p_T$ dependence is consistent within errors. As in Fig. 12, the error bars represent the statistical and systematic errors added in quadrature. The systematic errors dominate. $A(p_T,s)$ units are arbitrary since the cross section is self-normalized.

From the Tables IIa and IIb we conclude that the $p_T$ dependences are approximately the same in the two fragmentation cases (the values of $n_{pT}$ are within 5% of each other), whereas the D terms of the (1-$x_R$) exponent parameterizations by Eq. 6 <u>depend strongly</u> on the type of fragmentation side. In fact, such a D value for the Pb fragmentation side would correspond to inclusive jets at ≈ 10 TeV in p-p collisions implying that the quenching of jets observed in heavy ion collisions is also operative in p-p collisions, but at higher energies. This interpretation suggests an equivalency between the formation and quenching of jets at lower energies in A-A collisions with jet production at higher energies in p-p collisions and could be systematically studied by performing this analysis for jets produced in A-A collisions as a function of centrality.

### D. Comparison with Inclusive Jet Simulations

It is not the object of this paper to appraise the quality of the pQCD simulations of inclusive jet production, but it is of interest to check that the simulations show the same power law behaviors. Of the data examined in this work, from the CDF inclusive jets at 1.96 TeV published in 2009 to ATLAS inclusive jets at 13 TeV published in 2016 there is good agreement with simulations. The CDF analysis used the Midpoint jet clustering algorithm with a cone size R = 0.7 and proton and antiproton PDFs from [26] in conjunction with PYTHIA 6.2 [27]. The ATLAS collaboration used an anti-$k_T$ clustering algorithm with R = 0.4 and a more refined PDF set in Pythia 8.186 [28], [29].

As a demonstration of the agreement, the MC simulation SHERPA [30] has been compared with the 7 TeV ATLAS data, where the MC "data" were analyzed in the same way as the ATLAS 7 TeV inclusive jet data using the radial scaling formulation. The comparison of the ratios (MC/Data) of the respective fit parameters are given in the table below:

SHERPA-DATA COMPARISON: TABLE III

| Parameter | Ratio (SHERPA/Data) |
|---|---|
| $\alpha$ | $1.2 \pm 0.3$ |
| $n_{pT}$ | $0.98 \pm 0.02$ |
| D | $0.7 \pm 0.1$ |
| $n_{xR}0$ | $1.06 \pm 0.09$ |

Table III: The ratio of the fit parameters of the SHERPA simulation of the 7 TeV ATLAS inclusive jet data are given.

All parameters of the data – SHERPA comparison are consistent with each other, with the exception of D, which is smaller by about 30% in the SHERPA simulation ($3\sigma$). However, we note that his comparison of Data vs. MC using the $p_T$ and $x_R$ variables is quite sensitive to y-dependence and may be a useful test of data/MC in the future.

## III. SINGLE PARTICLE INCLUSIVE DATA

Since we find that the inclusive jet production at the LHC in p-p, p-Pb collisions and in $\bar{p}p$ collisions at the FNAL collider have power law dependences in both $p_T$ and $(1-x_R)$, it is interesting to analyze single hadron and prompt photon production.

### a. Prompt Photon Production

In prompt photon production, the photon is believed to come directly from the primordial hard parton scattering such as q g → q γ and higher order processes, such as the fragmentation process q g → q g γ. Unlike jet production, prompt photon production has no final state interaction other than the radiative fragmentation process above. Hence, the $E_T$ dependence as well as the $(1-x_R)$ dependence are important measures of the production mechanism without the influence of the final state processes.

A number of authors (for example see [31]) have extensively analyzed direct photon production in p-p collisions as a means to determine the nucleon gluon distribution, but not with our variables ($p_T$, $x_R$). For this study, we consider the prompt photon data determined by CMS at 7 TeV [32] and that of ATLAS at 8 TeV [33] and 13 TeV [34]. The photon data are analyzed in the same manner as the inclusive jet data – namely we compute the invariant cross sections d²σ/(E_T dE_T dη) and plot the results as a function of (1-x_R) for fixed E_T in order to determine A(E_T, s) and the (1-x_R) power law indices. The outcomes of the analysis for all three data sets are shown in Figs. 14 and 15.

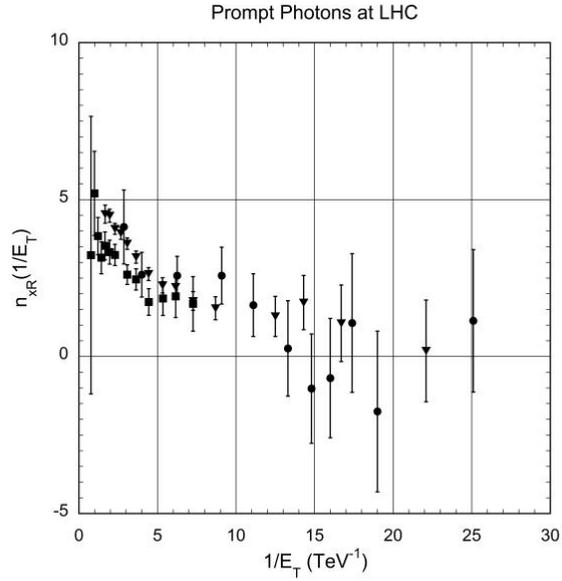

Fig. 14: Shown are the (1-x_R) exponents as a function of $1/E_T$ of the analyses of the 7 TeV CMS prompt photon data (closed circles) and the 8 TeV (triangles) and 13 TeV (squares) data sets of ATLAS. The highest point in both the 7 TeV CMS data and 8 TeV ATLAS data are weighted averages of the points where the errors are larger than 100% plotted at the weighted $1/E_T$ value. The errors bars are the quadrature sums of statistical and systematic contributions.

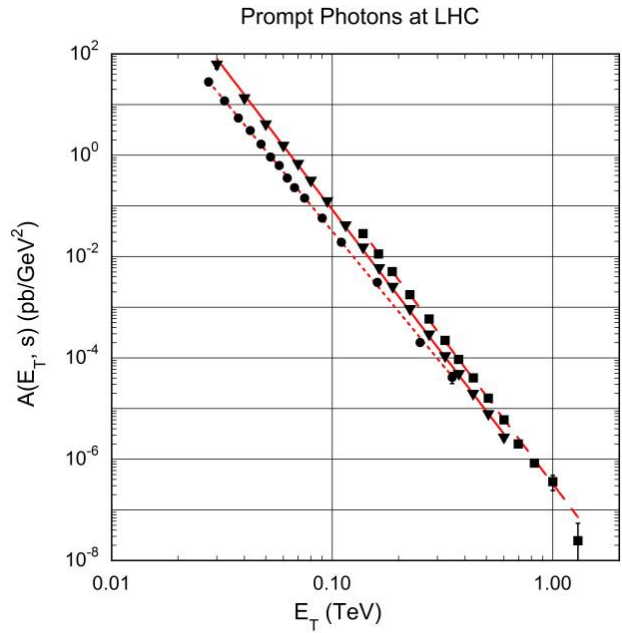

Fig. 15: The function A($E_T$, s) is plotted for the 7 TeV CMS prompt photon data (circles) and the 8 TeV and 13 TeV data of ATLAS (triangles and squares, respectively). The power law fits given in Table IVb are shown in red dotted line for 7 TeV CMS, solid and dashed lines for 8 TeV and 13 TeV ATLAS measurements, respectively.

As in the analysis of the inclusive jet data, we have determined the power indices $n_{xR}$ of Fig. 14 by fits of the function of Eq. 6, where $E_T$ replaced $p_T$ and the function A($E_T$,s) in Fig. 15 with Eq. 7. The results are given in Tables IVa,b. It is interesting to note in Table IVa that the parameter D for prompt photons is <u>negative</u> and appears to grow more negative with increasing √s. This is in contrast with the behavior for inclusive jets at the LHC where D is positive and increases with increasing √s. The parameter $n_{xR}0$ has an average value $<n_{xR}0> = 4.2 \pm 0.4$ which does not show a systematic energy dependence, although the dispersion of the data is large.

$N_{XR}$ FITS DIRECT PHOTON: TABLE IVa

| √s (TeV) | D (TeV$^{-1}$) | $n_{xR0}$ | $\chi^2$/d.f. | d.f. |
|---|---|---|---|---|
| 7 TeV CMS | $-0.20 \pm 0.07$ | $3.8 \pm 0.7$ | 0.4 | 13 |
| 8 TeV ATLAS | $-0.35 \pm 0.03$ | $4.7 \pm 0.1$ | 2.5 | 16 |
| 13 TeV ATLAS | $-0.43 \pm 0.09$ | $4.1 \pm 0.3$ | 0.4 | 12 |

Table IVa: The prompt photon invariant cross section parameters of the fits of the form of Eq. 6 of the power law indices of the $(1-x_R)^{n_{xR}}$ for constant $E_T$ are tabulated.

In Table IVb, where values of $\alpha$ and $n_{ET}$ are given, we see that the overall direct photon cross section grows with increasing √s as indicated by the fitted values of the $\alpha$-parameter. This is the same general behavior observed in our analysis of inclusive jet production cross sections.

$P_T$ FITS DIRECT PHOTON: TABLE IVb

| √s (TeV) | $\alpha$ (pb/GeV$^2$) TeV$^{n_{ET}}$ | $n_{ET}$ | $\chi^2$/d.f. | d.f. |
|---|---|---|---|---|
| 7 TeV CMS | $(1.7 \pm 0.2) \times 10^{-7}$ | $5.28 \pm 0.05$ | 0.7 | 13 |
| 8 TeV ATLAS | $(1.72 \pm 0.05) \times 10^{-7}$ | $5.69 \pm 0.01$ | 2.8 | 16 |
| 13 TeV ATLAS | $(3.3 \pm 0.1) \times 10^{-7}$ | $5.76 \pm 0.03$ | 1.4 | 12 |

Table IVb: The prompt photon invariant cross section parameters of the power law fits to A($E_T$, s) according to Eq. 7 are tabulated.

It is notable that the power law index $n_{ET}$ is less than the corresponding value for inclusive jets. Averaging over the three measurements in Table IVb (7 to 13 TeV) we find $\langle n_{ET} \rangle = 5.6 \pm 0.2$, whereas the average of the corresponding parameter for inclusive jets in the energy range 7 to 13 TeV is $\langle n_{pT} \rangle = 6.4 \pm 0.2$. This suggests that the prompt photon leaves the 'scene' of the primordial collision unencumbered; whereas jets, must tear themselves free of the QCD color fields. If we assume that $E_T$ dependence of $A(E_T, s)$ for prompt photons is a measure of the primordial hard parton scattering, then the fragmentation and hadronization operative in the production of jets in p-p and p-$\bar{p}$ collisions contribute to the jet and hadron $p_T$ dependence by roughly one more term ~ $1/p_T$ (a 3σ difference with these data).

## b. Hadron Production

A host of other inclusive production data were analyzed in the same way. Since the data extend to lower transverse momenta, we expect the parton intrinsic transverse momentum ($k_T$) to be an influence as well as transverse mass effects for heavy particle production. Moreover, for the case of charm production at the LHC we might expect a similar term that could arise from a production mechanism from the decay of a heavier "parent" particle. Hence, we fit the $p_T$ dependence with the form given in Eq. 8 below:

$$A(pT) = \frac{\alpha}{\left(\Lambda^2 + p_T^2\right)^{\frac{n_{pT}}{2}}} \qquad (8)$$

where the values of $\Lambda$, $\alpha$ and $n_{pT}$ are determined by a minimum $\chi^2$ fit [35]. A typical result is shown in Fig. 16 for $\pi^+$ using the compilation of reference [5].

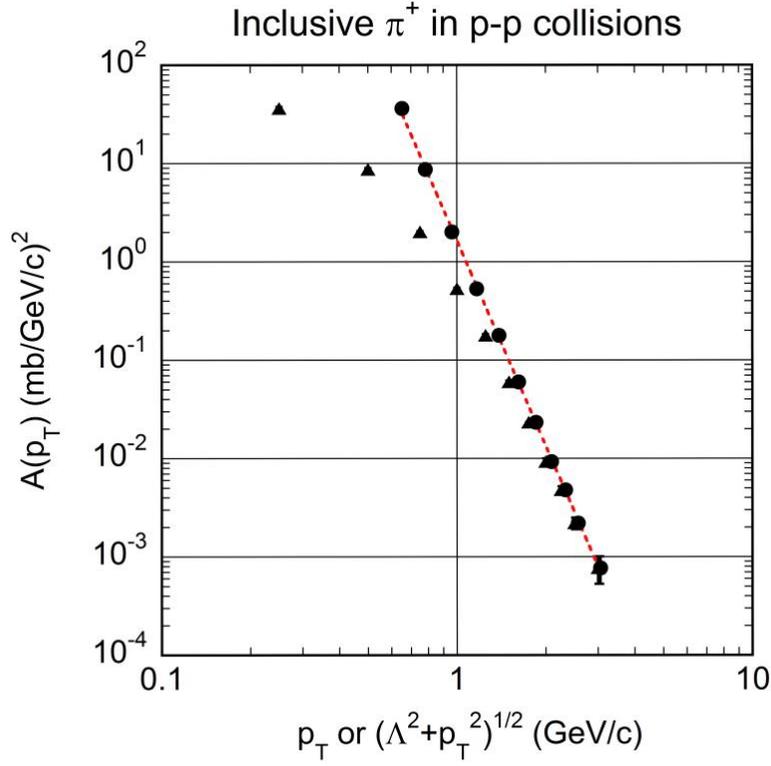

Figure 16: The inclusive $\pi^+$ data [5] are plotted with respect to the measured $p_T$ (black triangles) and with respect to $(\Lambda^2 + p_T^2)^{1/2}$ closed black circles. The red dotted line is the power law fit corresponding to $np_T = 6.94 \pm 0.04$ as given in Table V. A minimum $\chi^2$ fit of $\Lambda$ from Eq. 8 yields minimum at $\Lambda = 0.602 \pm 0.012$ GeV consistent with intrinsic parton $k_T$ of the nucleon [36].

The results of the $p_T$ power law fits are listed in Table V. We note that all the processes considered [36], including the five direct photon measurements have an average $p_T$ power-law dependence $A(p_T,s) \sim 1/p_T^{np_T}$ with an index $np_T \approx 6.1 \pm 0.6$. The inclusive production of light hadrons up to $K^+$ have a $\Lambda$ value consistent with the parton intrinsic $k_T$ ~ 0.6 GeV [37], whereas the J/$\psi$ and $\psi$(2S) production are consistent with a larger $\Lambda$ value ($\Lambda \geq 3.6$ GeV) which must provide an important clue about their production mechanism [38].

# $P_T$ FITS SINGLE PARTICLE INCLUSIVE PRODUCTION: TABLE V

| Index | Single Particle Inclusive Process | √s (TeV) | Λ(GeV) | σ(Λ) | $np_T$ | σ($np_T$) | <Λ> (GeV) | <σ(Λ)> |
|---|---|---|---|---|---|---|---|---|
| 1 | UA1 Direct γ | 0.546 | | | 5.7 | 0.3 | | |
| 2 | UA1 Direct γ | 0.63 | | | 5.9 | 0.5 | | |
| 3 | CMS Direct γ | 7 | | | 5.28 | 0.05 | | |
| 4 | ATLAS Direct γ | 8 | | | 5.69 | 0.01 | | |
| 5 | ATLAS Direct γ | 13 | | | 5.76 | 0.03 | | |
| 6 | $\pi^0$ 10 GeV to 63 GeV | 0.063 | 0.653 | 0.001 | 7.2 | 0.1 | | |
| 7 | ALICE $\pi^0$ $p_T \geq 0.5$ GeV | 2.76 | 0.8 | 0.2 | 6.1 | 0.3 | | |
| 8 | $\pi^+$ 10 GeV to 63 GeV | 0.063 | 0.60 | 0.02 | 6.9 | 0.1 | | |
| 9 | BRAHMS RHIC $\pi^+$ Ag-Ag | 0.062 | 0.56 | 0.07 | 5.7 | 0.5 | | |
| 10 | $\pi^-$ 10 GeV to 63 GeV | 0.063 | 0.607 | 0.004 | 6.86 | 0.02 | 0.77 | 0.09 |
| 11 | ALICE $\pi^\pm$ $p_T \geq 0.5$ GeV | 7 | 0.61 | 0.10 | 5.2 | 0.3 | | |
| 12 | $K^+$ 10 GeV to 63 GeV | 0.063 | 0.61 | 0.08 | 6.1 | 0.3 | | |
| 13 | $K^-$ 10 GeV to 63 GeV | 0.063 | 0.8 | 0.1 | 6.6 | 0.7 | | |
| 14 | ALICE $K^\pm$ $p_T \geq 0.5$ GeV | 7 | 0.94 | 0.10 | 5.5 | 0.3 | | |
| 15 | $p^-$ 10 GeV to 63 GeV | 0.063 | 0.9 | 0.1 | 6.8 | 0.5 | | |
| 16 | ALICE $p^\pm$ $p_T \geq 0.5$ GeV | 7 | 1.4 | 0.2 | 7.1 | 0.5 | | |
| 17 | LHCb D0 | 5 | 2.6 | 0.3 | 5.6 | 0.4 | | |
| 18 | LHCb D0 | 13 | 2.7 | 0.3 | 5.3 | 0.3 | | |
| 19 | LHCb $Ds^+$ | 5 | 2.5 | 0.8 | 5.3 | 0.8 | 2.80 | 0.25 |
| 20 | LHCb $Ds^+$ | 13 | 3.1 | 0.8 | 5.6 | 0.7 | | |
| 21 | LHCb $D*^+$ | 5 | 2.8 | 0.7 | 5.9 | 0.9 | | |
| 22 | LHCb $D*^+$ | 13 | 3.1 | 0.7 | 5.5 | 0.6 | | |
| 23 | ATLAS: prompt J/ψ | 5.02 | 3.6 | 0.3 | 7.0 | 0.1 | | |
| 24 | ATLAS: prompt J/ψ | 7 | 2.7 | 1.6 | 6.6 | 0.2 | | |
| 25 | CMS: prompt J/ψ | 7 | | | 6.7 | 0.04 | 3.6 | 0.5 |
| 26 | ATLAS: prompt J/ψ | 8 | 3.0 | 1.7 | 6.4 | 0.2 | | |
| 27 | CMS: prompt J/ψ | 13 | | | 5.92 | 0.05 | | |
| 28 | LHCb: prompt J/ψ | 13 | 4.4 | 0.4 | 7.0 | 0.5 | | |
| 29 | ATLAS: prompt ψ(2S) | 7 | 4.1 | 2.5 | 6.6 | 0.5 | 4.3 | 2.0 |
| 30 | ATLAS: prompt ψ(2S) | 8 | 4.5 | 1.5 | 6.6 | 0.2 | | |
| 31 | ATLAS: non-prompt J/ψ | 5.02 | 7.1 | 1.2 | 6.5 | 0.4 | | |
| 32 | ATLAS: non-prompt J/ψ | 7 | 5.8 | 1.6 | 6.1 | 0.3 | 6.2 | 1.0 |
| 33 | ATLAS: non-prompt J/ψ | 8 | 7.4 | 0.7 | 6.1 | 0.1 | | |
| 34 | LHCb: non-prompt J/ψ | 13 | 4.6 | 0.3 | 5.7 | 0.3 | | |
| 35 | ATLAS: non-prompt ψ(2S) | 7 | 4.1 | 2.8 | 5.6 | 0.4 | 4.8 | 2.3 |
| 36 | ATLAS: non-prompt ψ(2S) | 8 | 5.4 | 1.7 | 5.7 | 0.3 | | |
| 37 | LHCb B0 | 7 | 6.5 | 2.2 | 5.5 | 1.2 | | |
| 38 | LHCb B± | 7 | 6.4 | 1.1 | 5.5 | 0.6 | 6.7 | 0.3 |
| 39 | LHCb Bs0 | 7 | 7.1 | 2.2 | 5.9 | 1.3 | | |
| | | | | <$np_T$> | 6.1 | 0.6 | | |

Table V: Tabulated are the values of the power law fits to various processes through Eq. 8. All processes are for inclusive production in p-p collisions except for Ag-Ag collisions of index 9. For those entries of the table where √s = 0.063 TeV the tabulated √s value is the maximum of the data set, which also includes lower √s values down to 10 GeV in some entries [5]. The values of Λ and associated errors are determined by the curvature $\chi^2$ function about its minimum. The 7 TeV CMS prompt J/ψ data are consistent with Λ=0, unlike the other measurements, but with a large error and for this reason Λ and σ(Λ) for this entry are left blank. Entries1 through 5 for direct γ have a minimum $E_T \gg k_T \sim 0.6$ GeV and thus no sensitivity to the $k_T$ (Λ) value. Each table index is referenced [36].

The power law indices average = 6.1 ± 0.6. The values of Λ for the single light ($\pi^{\pm,0}$, $K^+$) particle inclusive production are consistent with the intrinsic $k_T$, whereas the values of Λ for heavier particles, such as direct and indirect J/ψ production, are strongly influenced by secondary decay chains.

## IV. LINE COUNTING, HIGHER TWISTS, DIQUARKS

Using the radial scaling formulation discussed above and examining Tables Ib, IIb and V, it is remarkable that the $p_T$ factorized part of the invariant cross sections is a power law with the behavior $A(p_T,s) \approx \alpha(s)/p_T^6$ with essentially all the s-dependence confined in the term $\alpha(s)$. This is true for inclusive jet production in high-energy p-p, $\bar{p}$-p collisions and inclusive single particle production in p-p and inclusive $\pi^+$ production in Ag-Ag collisions. (Direct photon production favors ~ $1/p_T^{5.7 \pm 0.2}$.) Further, the exponent $n_{xR}$ of $(1-x_R)$ for LHC inclusive jet production is found to be a linear function of $1/p_T$ with the slope parameter D increasing with increasing √s. Inclusive jet production in p-Pb collisions shows the same behavior but has a significantly different D-value depending on the fragmentation side (proton-forward or the Pb-forward).

It is well known that the dimensions of the invariant cross section are dependent on the number of active fields that hard-scatter to produce the detected jet or the particle in nucleon-nucleon scattering. By this argument the matrix element for the hard-scattering M ~ (Mass)$^{4-nA}$, where nA is the number of active fields that scatter [39], [40], [41], [42], [43]. Since the invariant cross section has the form given by Eq. 9 we would expect the $p_T$ dependence of the invariant cross section by this argument to be given by Eq. 10.

$$\frac{d^2\sigma}{p_T dp_T dy} \propto \frac{|M|^2}{\hat{s}^2} \qquad (9)$$

$$\frac{d^2\sigma}{p_T dp_T dy} \propto \frac{1}{p_T^{2n_A - 4}} \qquad (10)$$

Referring to Fig. 17 below, we note, for example, that u-d elastic scattering, or g-u elastic scattering all involve four active fields and therefore would have a $p_T$ dependence given by Eq. 11a. Whereas, for example the diagram at the bottom-left part of the figure involves quark-quark scattering with two radiated gluons, one in the final state and one that forms a diquark. This diagram involves five fundamental fields and consequently would have a $p_T$ dependence given by Eq. 11b.

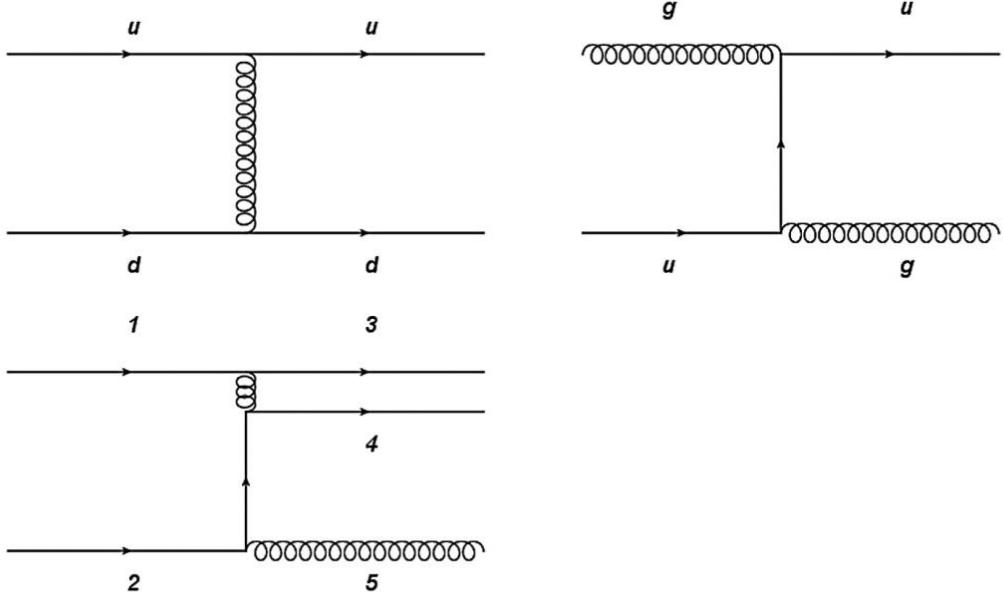

Fig. 17: Shown are Feynman diagrams for u-d quark elastic scattering, g-u elastic scattering and quark-quark scattering with two gluons, one radiated (line 5) to the final state. Note that lines 3 and 4 denote a diquark. where the number of active lines $n_A = 5$. (After Arleo, Moriond QCD 2010 [41].)

$$\frac{d^2\sigma}{p_T dp_T dy} \propto \frac{1}{p_T^4} \quad (11a)$$

$$\frac{d^2\sigma}{p_T dp_T dy} \propto \frac{1}{p_T^6} \quad (11b)$$

It is noteworthy that all the processes tabulated above seem to favor the $p_T$ dependence given in Eq. 11b over a wide range of energies, rather than the lowest order scattering which has a $p_T$ dependence given by Eq. 11a. Other diagrams, such as ones with a radiated gluon from a scattered quark, would correspond to a 2 → 3 scattering resulting in a higher $p_T$ power. The $p_T$ power-law index ~ 6 is a surprise since one would expect an index of ~ 4 for parton-parton (2 → 2) hard elastic scattering at lowest order.

A number of authors have observed ([44], [45], [46], [47], [48], [49]) that the effective $p_T$-power is larger than the expected dimensional limit of $2 \rightarrow 2$ scattering but some researchers find that the $p_T$ power seems to depend on the process. Some of these analyses explore the limit to scaling as a function of $x_T \rightarrow 0$, which we have shown does not respect the kinematic boundary and therefore mixes the kinematic boundary suppression with the underlying $p_T$-dependence. An appraisal of one of these studies [40] is given in Appendix A. On the contrary, we find that the average $p_T$ power is $<n_{pT}> = 6.1 \pm 0.6$ for all 46 inclusive single photon/hadron/jet data sets considered in p-p and $\bar{p}$-p collisions (for inclusive jets in p-p and $\bar{p}$-p $<n_{pT}> = 6.5 \pm 0.3$ and single particle inclusive cross sections $<n_{pT}> = 6.1 \pm 0.6$). The invariant cross section dimensional limit $n_{pT} = 4$ is therefore disfavored by 3.8 $\sigma$. Even direct $\gamma$ production disfavors $n_{pT} = 4$ by 8 $\sigma$ ($<n_{ET}> = 5.6 \pm 0.2$ vs. 4). This behavior is consistent with a dominant $2 \rightarrow 3$ hard scatting that is saturated at a relatively low $\sqrt{s}$ – thereby becoming independent of process and COM energy (See Feynman, Field and Fox [2]).

The dominant $2 \rightarrow 3$ scattering is consistent with an intrinsic diquark inside the nucleon, although a $2 \rightarrow 2$ scattering with a radiated gluon from one of the final quark legs would also have a cross section of the same $p_T$-dimension. Evidence of diquark correlations in the proton have been discussed for some time [50]. Recently, data from JLab [51] supports the notion of diquarks affecting the proton elastic form factors. Lattice QCD calculation also indicate that there is a strong association of the u-d quarks in the proton that forms a singlet (diquark) state [52], [53].

## V. $A(p_T,s)$ FOR JETS AS A QUADRATIC IN $\ln(p_T)$

We have noted in Tables Ib and IIb that the $p_T$ power-law fits for inclusive jets had rather unlikely $\chi^2$ values. A close examination of the fit-data relation reveals a systematic deviation from a pure power law – namely, there is a small curvature making the $p_T$-dependence less steep at low $p_T$ than at high $p_T$. The effect is illustrated in Fig. 18 where we plot the residuals of the power law fit of the 13 TeV ATLAS inclusive jet data as a function of $\ln(p_T)$. In order to make the discrepancy clear, the baseline power law fit was determined by treating all errors the same. The error bars in the figure represent the statistical and systematic errors added in quadrature as in Fig. 6.

We notice that the residuals of the single power-law fit plotted in Fig. 18 can be quite well fitted to a quadratic in $\ln(p_T)$. This suggests that the $p_T$-dependence of the invariant cross section for inclusive jets at 13 TeV is a function of the type:

$$\ln(A(p_T,s)) = \beta(s)\ln(p_T)^2 - n_{p_T}\ln(p_T) + \rho(s) \tag{12}$$

An equivalent form of Eq. 12 makes evident the underlying $p_T$ power law with a moderating term controlled by the parameter β and is given by:

$$A(p_T,s) = \exp\left(\beta(s)(\ln(p_T))^2\right)\frac{\alpha(s)}{p_T^{n_{p_T}}} \tag{13}$$

where $\alpha(s) = \exp(\rho(s))$.

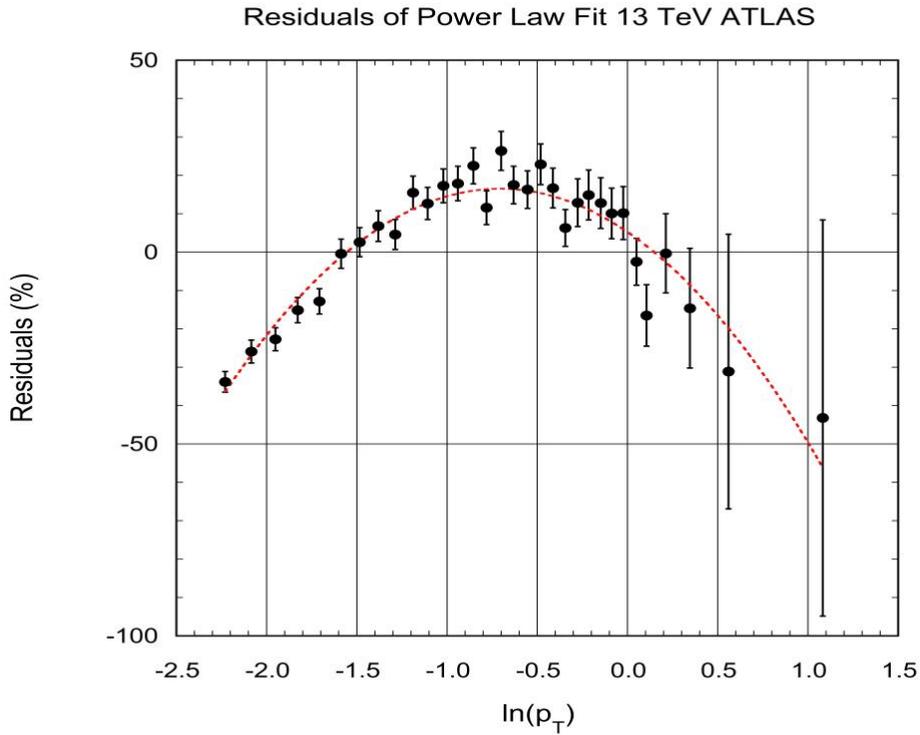

Fig. 18: The residuals of the power law fit to 13 TeV ATLAS inclusive jet data. A simple power law ~ $1/p_T^{6.45 \pm 0.037}$ fits the data ($\chi^2$/d.f. = 8.2 for 30 d.f.). The residuals are confined to be within ~ ± 30% over 9 orders of magnitude in $p_T$. The dotted red line is a quadratic fit in $\ln(p_T)$ to the residuals.

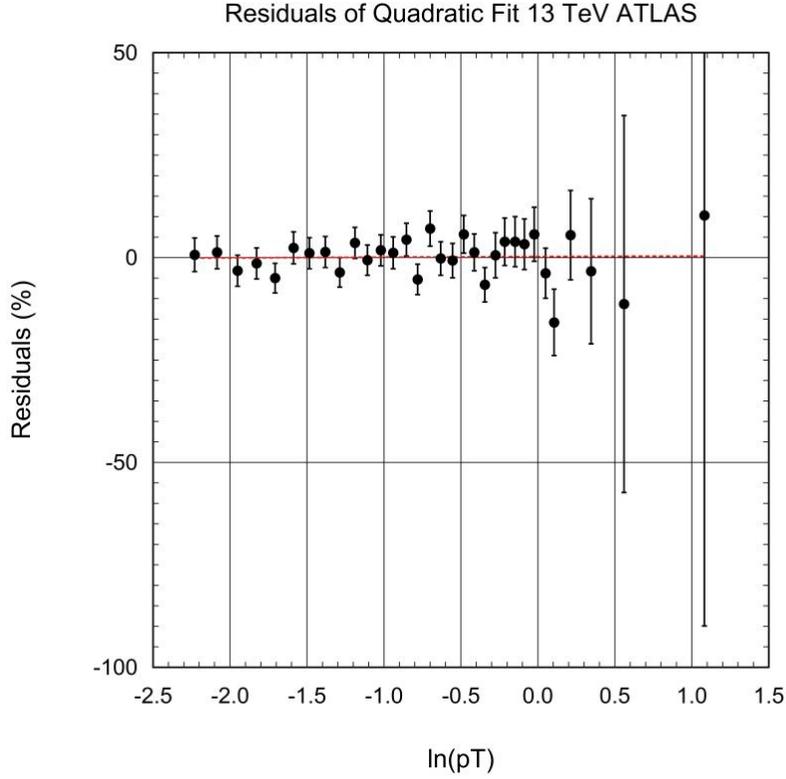

Fig. 19: The residuals of the quadratic $\ln(p_T)$ fit to 13 TeV ATLAS inclusive jet data. Adding the quadratic term in $\ln(p_T)$ improves the $\chi^2$ of the fit and reduces the residuals to no further visible dependence on $p_T$ ($\chi^2$/d.f. = 0.8 for 29 d.f.).

Fitting $A(p_T,s)$ of ATLAS and CMS inclusive jets and the inclusive jets of CDF and D0 1.96 TeV to Eq. 12, we find the parameter values given in Table VI. The residuals of the ATLAS 13 TeV inclusive jets of this quadratic $\ln(p_T)$ fit are shown in Fig. 19.

QUADRATIC LN($P_T$) FITS: TABLE VI

| $\sqrt{s}$ (TeV) | $\beta$ | $\alpha$ (pb/GeV$^2$) TeV$^{n_{pT}}$ | $n_{pT}$ | $\chi^2$ /d.f. | d.f. |
|---|---|---|---|---|---|
| 1.96 $\bar{p}$-p CDF 1.96 $\bar{p}$-p D0 | 0.03 ± 0.2 | (1.6 ± 0.8) x 10$^{-6}$ | 6.7 ± 0.6 | 0.92 | 38 |
| 2.76 p-p ATLAS | -0.23 ± 0.09 | (1.3 ± 0.6) x 10$^{-6}$ | 7.5 ± 0.4 | 1.17 | 7 |
| 7 p-p ATLAS | -0.38 ± 0.05 | (1.0 ± 0.1) x 10$^{-5}$ | 7.8 ± 0.2 | 2.50 | 13 |
| 8 p-p CMS | -0.38 ± 0.02 | (2.1 ± 0.1) x 10$^{-5}$ | 7.62 ± 0.05 | 4.3 | 32 |
| 13 p-p ATLAS | -0.26 ± 0.01 | (9.2 ± 0.1) x10$^{-5}$ | 6.92 ± 0.02 | 0.77 | 29 |
| 13 p-p CMS | -0.32 ± 0.04 | (8.7 ± 0.2) x10$^{-5}$ | 7.03 ± 0.07 | 0.48 | 26 |

Table VI: The parameters of the quadratic fit for inclusive jets in $\ln(p_T)$ defined by Eq. 13 are tabulated for p-p scattering and $\bar{p}$-p of CDF and D0 statistically combined. The fit parameters were determined with $p_T$ values in TeV. Correlations between parameters have been neglected.

The 7 TeV MC simulation (SHERPA [30]) was analyzed in the same manner yielding $\beta =$ $-0.2 \pm 0.06$, $\alpha = 1.5 \pm 0.2 \times 10^{-5}$ (pb/GeV$^2$) TeV$^{n_{pT}}$ and $n_{pT} = 7.1 \pm 0.2$.

Adding the $\beta$-term of Eqs. 12, 13 improves the $\chi^2$ of the A($p_T$,s) fits for inclusive jet production quite significantly as noted in Table VI compared with Table Ib and seen in Fig. 19. However, the data are good enough to draw only rough conclusions about the systematics of the s-dependence. The $\beta$- and $n_{pT}$- terms are roughly independent of $\sqrt{s}$, whereas the $\alpha$-term grows roughly linearly with increasing s.

While we have interpreted the deviations from a pure $p_T$-power law as 'real', an uncorrected nonlinearity in the jet energy calibration could also be contributing. The power law index is independent of an overall energy scale calibration which would only contribute an additive term in the linear fits to ln($p_T$) not affecting the value of $n_{pT}$. However, both the power law index, $n_{pT}$ as well as a $\beta$-term in Eq. 12 above would be affected by a calorimetry nonlinearity. We note that the form of Eq. 12 is consistent with a log-normal distribution and also with a Sudakov-like form factor [54] with suitable choice of parameters.

## VI. SUMMARY

This paper is an attempt to characterize inclusive jet production from p-p, p-Pb and $\bar{p}$-p collisions at high energy in minimal common terms and to compare these reactions with single particle inclusive reactions – including charm production and direct photon production at the LHC. Analyzing the invariant cross sections for inclusive jets, single hadrons and prompt photons in p-p and $\bar{p}$-p collisions reveals a simple structure – namely that the invariant cross sections factorizes into a product of two power laws – one in $p_T$ and the other in (1-$x_R$). All these inclusive invariant cross sections are of the form given in the equation below:

$$\frac{d^2\sigma}{p_T dp_T dy}(s, p_T, y; \alpha, \Lambda, n_{pT}, m, D, n_{xR0})$$

$$= \frac{\alpha(s)}{\left(\Lambda^2 + p_T^2\right)^{\frac{n_{pT}}{2}}} \left(1 - \frac{2\sqrt{\left(p_T^2 \cosh^2(y)(1+(m^2/p_T^2)\tanh^2(y)) + m^2\right)}}{\sqrt{s}}\right)^{\frac{D(s)}{p_T}+n_{xR0}} \quad (14)$$

$$= \frac{\alpha(s)}{\left(\Lambda^2 + p_T^2\right)^{\frac{n_{pT}}{2}}} (1-x_R)^{\frac{D(s)}{p_T}+n_{xR0}}$$

,

where the kinematic variables are s, $p_T$ and y and the parameters are α, Λ, $n_{pT}$, m (or $m_J$), D and $n_{xR0}$ described in the text. It is interesting to note that the s-dependence for fixed $x_R$ is confined to the parameters α(s) and D(s), which grow linearly with increasing s. At high $p_T$ ($p_T \gg$ m) Eq. 14 simplifies to:

$$\frac{d^2\sigma}{p_T dp_T dy}(s, p_T, y; \alpha, n_{pT}, D, n_{xR0}) = \frac{\alpha(s)}{p_T^{n_{pT}}}\left(1 - \frac{2p_T \cosh(\eta)}{\sqrt{s}}\right)^{\frac{D(s)}{p_T}+n_{xR0}}$$

$$= \frac{\alpha(s)}{p_T^{n_{pT}}}(1-x_R)^{\frac{D(s)}{p_T}+n_{xR0}} \quad (15)$$

.

The $p_T$-power laws of Eqs. 14, 15 are uncovered by using the $x_R$ variable to extrapolate the invariant cross sections at various constant $p_T$ values as a function of (1-$x_R$) to the limit $x_R \to 0$ at fixed $\sqrt{s}$. This procedure determines the underlying A($p_T$,s) ≈ α(s)/$p_T^{n_{pT}}$ function independent of $x_R$. All the processes analyzed in this paper have a power law index confined to 5.3 < $n_{pT}$ < 7.1. In broad terms, the $p_T$ powers of inclusive cross sections are roughly independent of √s and process (see appendix). By averaging all data analyzed (jets, photons, hadrons) the naive dimensional limit of the invariant cross section $n_{pT}$ = 4 is disfavored by 3.8 σ. (An even stronger exclusion of $n_{pT}$ = 4 is obtained by considering the weighted average <$n_{pT}$> = 6.296 ± 0.005). The data analyzed are consistent with five interacting partons in a 2 → 3 primordial hard scattering that is also a signature of an emergent diquark in the nucleon and that of 2 → 2 scattering with a gluon radiated in one of the final quark lines. A closer examination shows that A($p_T$,s) only roughly follows a simple power law in $p_T$ for inclusive jets at the LHC. In this case, the $p_T$ function is much better fit with a log-normal distribution, or equivalently a power law modified by a Sudakov-like form factor (FF ~ exp[$\beta(s) \ln^2(p_T)$]).

Our procedure involves analyzing the invariant cross sections for a <u>fixed value of $\sqrt{s}$</u> in order to determine the $p_T$ and the $x_R$ dependences. Since the s-dependence of the $p_T$ and $x_R$ dependences have to be estimated by comparing the analysis of different values of $\sqrt{s}$, it is therefore mandatory to have data sets at several values ($\geq 3$) of y ($\eta$, or $\theta$) as well as several values of $p_T$ and $\sqrt{s}$ in order to separate the $p_T$, $x_R$ and $\sqrt{s}$ dependencies.

The s-dependence of our jet fit parameters, shown in Fig. 11 and in Tables Ia and Ib is an indication that jets at low $p_T$ and high $\sqrt{s}$ will be strongly quenched. This may be an important factor in planning experiments at a 100 TeV p-p collider.

Without a detailed analysis of the various experimental systematic errors, which is beyond the scope of this work, it is not clear in some cases whether the relatively small differences in parameter values seen are evidence of real differences, such as the different of $p_T$-powers of prompt photon production from inclusive jets, or uncorrected systematic effects. Better data and more sophisticated analyses, which for example would involve corrections for finite $p_T$ and $x_R$ bins, will help resolve these issues. In fact, an examination of the fit values by comparing the parameters for ATLAS and CMS inclusive jets of $n_{pT}$ for inclusive jets in Table Ib, indicates that $\Delta(n_{pT}) \sim 0.4$ is within the systematic errors of this analysis.

One aspect of this analysis not explored in detail is the $x_R$-dependences. Unlike the $p_T$ behavior, the $x_R$ side is process, as well as s-dependent through the D-term and is therefore rich phenomenologically. In quark-line counting schemes the exponent of $(1-x_R)$ is dependent on the number of spectator fields and is given by $2n_{spectator} -1$. Examining this feature of the inclusive charm cross section should offer important information about the production mechanism.

The inclusive jet and prompt photon invariant cross sections are well replicated by simulation. In fact, pQCD and various MC programs, such as Phythia [27], [28] throughout its historical development, show power laws in $p_T$ as well as in the variable $(1-x_R)$. Hence, the simple behavior revealed in this analysis is already deeply embedded in the simulations and therefore 'understood'. However, the simple factorized form of the invariant inclusive cross sections, as worked out by this analysis using the $x_R$ variable to control phase space, shows a simple structure that may be useful in uncovering non-trivial signatures independent of kinematic effects.

In the original formulation of radial scaling, it was posited that <u>all</u> the s-dependence of the inclusive invariant cross sections was in the scaling variable, $x_R \approx 2p_T\cosh(\eta)/\sqrt{s}$, and that the $p_T$ and $x_R$ dependences of the invariant cross sections completely factorized. This turned out to be not generally true. Data taken at higher collision energies showed that there is an additional s-dependence in the $\alpha(s)$ term, beyond the simple $x_R$ function, that arises from the QCD-evolution of the parton, fragmentation and hadronization functions. Moreover, we found in our analysis that the $(1-x_R)$ power index, $n_{xR}$, has a $p_T$– dependence that is controlled in our formulation by the 'D-term'. Thus, the simple factorization of the invariant inclusive cross sections into a $p_T$ - part and an $x_R$ - part is broken.

The $x_R$ variable, unlike $x_T$ or $x_\parallel$, has utility in that it quantifies the fraction of the energy of the jet or particle with respect to the kinematic limit in inclusive cross sections that is independent of angle in COM frame. Controlling this faction breaks the conflation of a purely kinematic effect from a deeper dynamical behavior that seems to have confused several authors. The approximate scaling variable, developed over 40 years ago in the analysis of inclusive particle production in p-p collisions, still finds utility in uncovering simple power laws in inclusive jets, photons and charm in both p-p and p-Pb collisions at the LHC. Now that the data from the LHC are reaching maturity in broad kinematic ranges, it will be interesting to analyze their broad trends using our formulation.

**ACKNOWLEDGEMENTS**
The author thanks his colleagues, especially S. Brodsky, D. Duke, P. Fisher, L. Rosenson and J. Thaler, for interesting discussions and suggestions.

detector", arXiv:1605.03495v2 [hep-ex] 12 Oct 2016, JHEP 06 005 (2016), DOI: 10.1007 / JHEP 08 005 (2016).

34. ATLAS Prompt Photons: The ATLAS Collaboration, "Measurement of the cross section for inclusive isolated-photon production in p p collisions at √s = 13 TeV using the ATLAS detector", arXiv:1701.06882v1 [hep-ex] (24 Jan 2017), CERN-EP-2016-291, 25th January 2017, submitted to Phys. Lett. B.

35. Typically the function LINEST: (https://support.office.com/en-us/article/LINEST-function-84D7D0D9-6E50-4101-977A-FA7ABF772B6D) was applied to the logs of the data in order to determine the power law indices. Having determined trial fit parameters in this way, the $\chi^2$ of the fit function vs. data was minimized to determine $\Lambda$, $\alpha$ and $n_{pT}$.

36. References for TABLE V: Index below refers to the first column of Table V.

| Index | Reference |
|---|---|
| 1,2 | UA1: "Direct Photon Production at the CERN Proton-Antiproton Collider", UA1 Collaboration (c. Albajar, et al.) Phys. Lett. B 209, 385-396 (4 Aug. 1988). |
| 3 | CMS: "Measurement of the Differential Cross Section for Isolated Prompt Photon Production in pp Collisions at 7 TeV", CMS Collaboration (S.Chatrchyan, et al.) ; Phys. Rev. D 84 (2011) 052011, http://inspirehep.net/record/922830/data. |
| 4 | ATLAS: "Measurement of the inclusive isolated prompt photon cross section in pp collisions at √s = 8 TeV with the ATLAS detector", The ATLAS Collaboration; JHEP 06 (2016) 005, DOI: 10.1007/ JHEP 08 (2016) 005, arXiv:1605.03495v2 [hep-ex] (12 Oct. 2016). |
| 5 | ATLAS: "Measurement of the cross section for inclusive isolated-photon production in pp collisions at √s = 13 TeV using the ATLAS detector", The ATLAS Collaboration; arXiv:1701.06882v1 [hep-ex] (24 Jan. 2017). |
| 6, 8, 10, 12, 13, 15 | F. E. Taylor, et al., "Analysis of radial scaling in single-particle inclusive reactions", Phys. Rev. D 14, 1217-1242 (1 September 1976), Erratum Phys. Rev. D 15, 3499 (1977). Data were analyzed from Table IV of this reference. |
| 7 | ALICE: "Neutral pion production at midrapidity in pp and Pb-Pb collisions at √sNN = 2.76 TeV", ALICE Collaboration; Eur. Phys. J. C 74 (2014) 3108, Data: Table 2 http://hepdata.cedar.ac.uk/view/ins1296306. |

# APPENDIX

Arleo, *et al.* [40] have analyzed a number of inclusive measurements, such as inclusive single particle production in p-p scattering and inclusive jet production at the SPS and FNAL collider. They find the $p_T$ power depends on the process as given in Fig. A1 and is strikingly different from our analysis which finds all processes examined to be clustered around $n^{exp} \sim 6.5$. Of particular note is the analysis of inclusive jets at CDF and D0 (triangles in the fig. below) where the exponent $n^{exp} \sim 4.5$ is found.

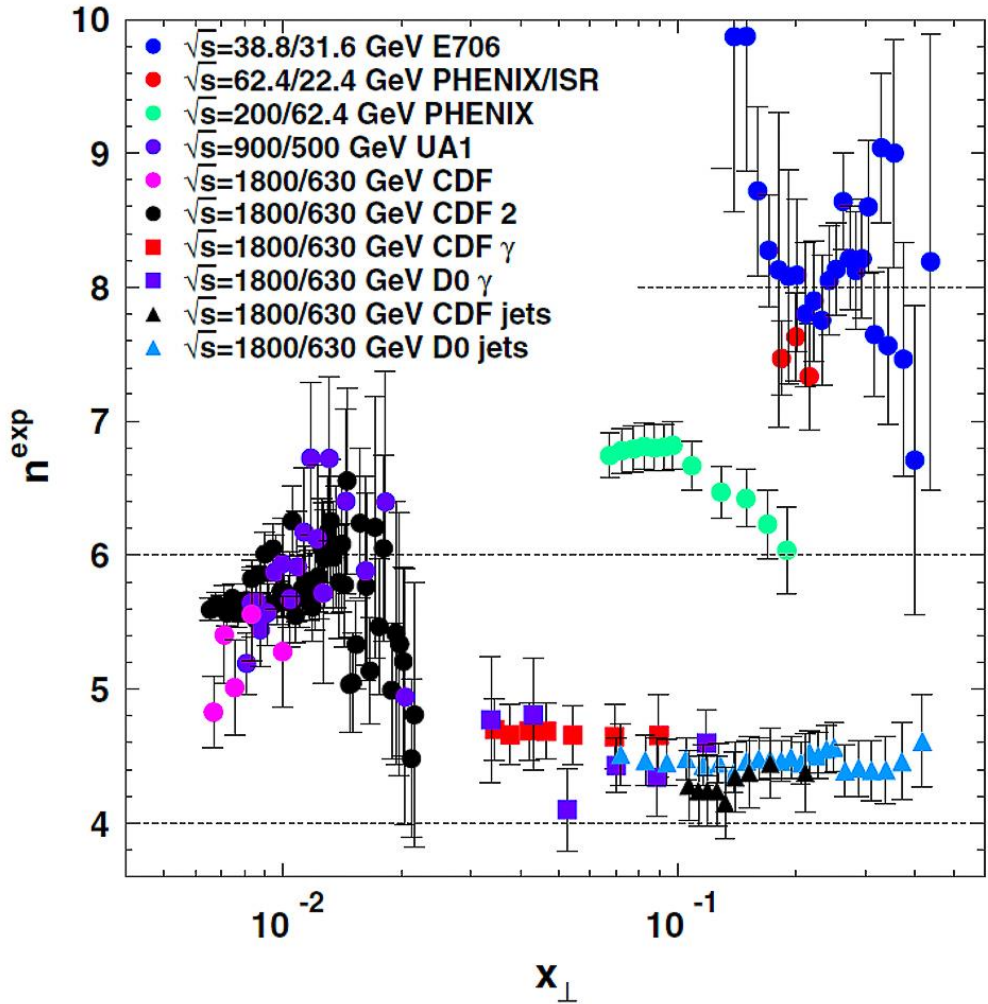

Fig. A1: The effective $p_T$ power $n^{exp}$ from the analysis of Arleo et al. [40] is shown for various processes. This result is strikingly different from this analysis, which finds $n_{pT}$ for all processes examined to be clustered around $n^{exp} \sim 6.4 \pm 0.5$. Note that the highest exponents determined by Arleo et al. are from comparisons of the lowest √s data.

Arleo et al. posit that the invariant cross sections depend on $p_T$ and $x_T$ defined by $x_T = 2p_T/\sqrt{s}$. By computing the ratios of cross sections at different values of $\sqrt{s}$ they are able to extract the effective $p_T$ power denoted by $n^{exp}$. In their analysis, the invariant cross section is given in Eq. A1:

$$\sigma^{inv} \equiv E\frac{d^3\sigma}{dp^3}(AB \to CX) = \frac{F(x_T, \theta)}{p_T^n} \tag{A1}$$

The analysis rests on the assumption that the function $F(x_T, \theta_1) \approx F(x_T)$ and that the s-dependence of the cross section is entirely through the $x_T = 2p_T/\sqrt{s}$ term so that invariant cross section can be written as:

$$\sigma^{inv}(AB \to CX) \propto \frac{(1-x_T)^{2n_{spectator}-1}}{p_T^{2n_{active}-4}} \tag{A2}$$

The $n^{exp}$ value is determined by ratio of the cross sections and the respective ratio of the COM energies. From Eqs. A1 and A2 using $p_T = x_T \sqrt{s}/2$:

$$n^{exp} = \frac{-\ln\left(\sigma^{inv}(x_T, \sqrt{s_1})/\sigma^{inv}(x_T, \sqrt{s_2})\right)}{\ln\left(\sqrt{s_1}/\sqrt{s_2}\right)} \tag{A3}$$

But by the radial scaling hypothesis, the form in Eq. A2 is generally not true since the function $F(x_T)$ is really a function of $p_T$, $\theta$ ($x_R$) and <u>importantly of $\sqrt{s}$ through the $\alpha(s)$-term.</u>

In order to show the flaw in this analysis at least for LHC inclusive jets, we take our parameterization of the LHC inclusive jets at $\sqrt{s}$ = 13 TeV and 2.76 TeV given in Table I to determine $n^{exp}$ in the same way. We compute the ratio of 13 TeV to 2.76 TeV ATLAS inclusive jets to examine the cross section ratio as $x_T \to 0$. The result is shown in Fig. A2 where we plot $n^{exp}$ given by Eq. A3 as a function of $x_T$ for various fixed jet COM angles $\theta$. We find that our evaluation of Eq. A1 yields an effective $p_T$ power of ~ 4 in the limit $x_T \to 0$ consistent with the analysis of Arleo et al. for CDF and D0 inclusive jets at 1.8 TeV / 0.63 TeV. Similar results are obtained when we compare 13 TeV jets to 7 TeV jets.

It is interesting to note that this result for the ATLAS data depends strongly on the s-dependence of the cross sections through the $\alpha$-term of Eq. 7. Setting the term $\alpha = 1$, but leaving the other parameters ($n_{pT}$, D and $nx_{R0}$) at their fit values one finds $n^{exp} \approx 6.3$ as $x_T \to 0$. Setting all parameters to the same value ($\alpha = 1$, $n_{pT} = 6.3$, D = 0 and $nx_{R0} = 3.5$) we find $n^{exp} \approx 6.3$. But putting in the measured s-dependence of $\alpha$ and leaving the other parameters of the cross section the same ($n_{pT} = 6.3$, D = 0 and $nx_{R0} = 3.5$) we find $n^{exp} \approx 4.4$. These results indicate that $n^{exp} \approx 4$ of Arleo, *et al.* is a result of the s-dependence of the cross section, which in our parameterization is mostly through the $\alpha$-term for small $x_R$, and not a true measure of the intrinsic $p_T$-dependence.

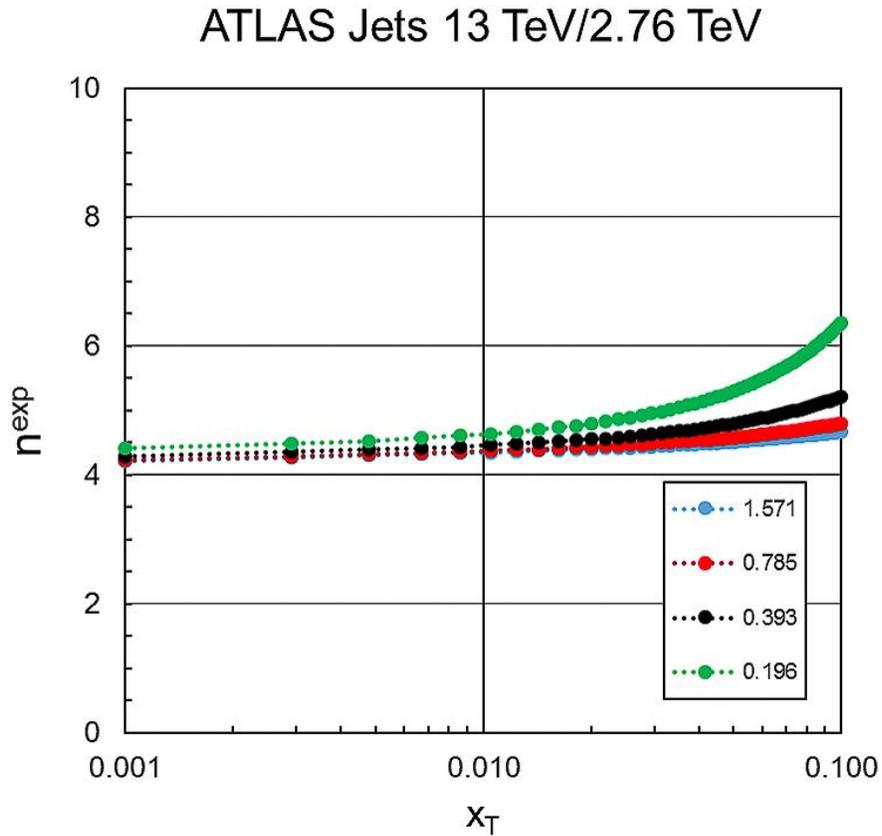

Fig. A2: The ATLAS inclusive jet cross section parameterizations given in Table I for √s = 2.76 TeV and 13 TeV are used to evaluate Eq. A1 above. The various lines are for fixed COM angles starting at $\theta = \pi/2$ down to 0.196 radians. All lines converge to $n^{exp} \sim 4.3$ even though the underlying $p_T$ dependence is $\sim 1/p_T^{6.3}$.

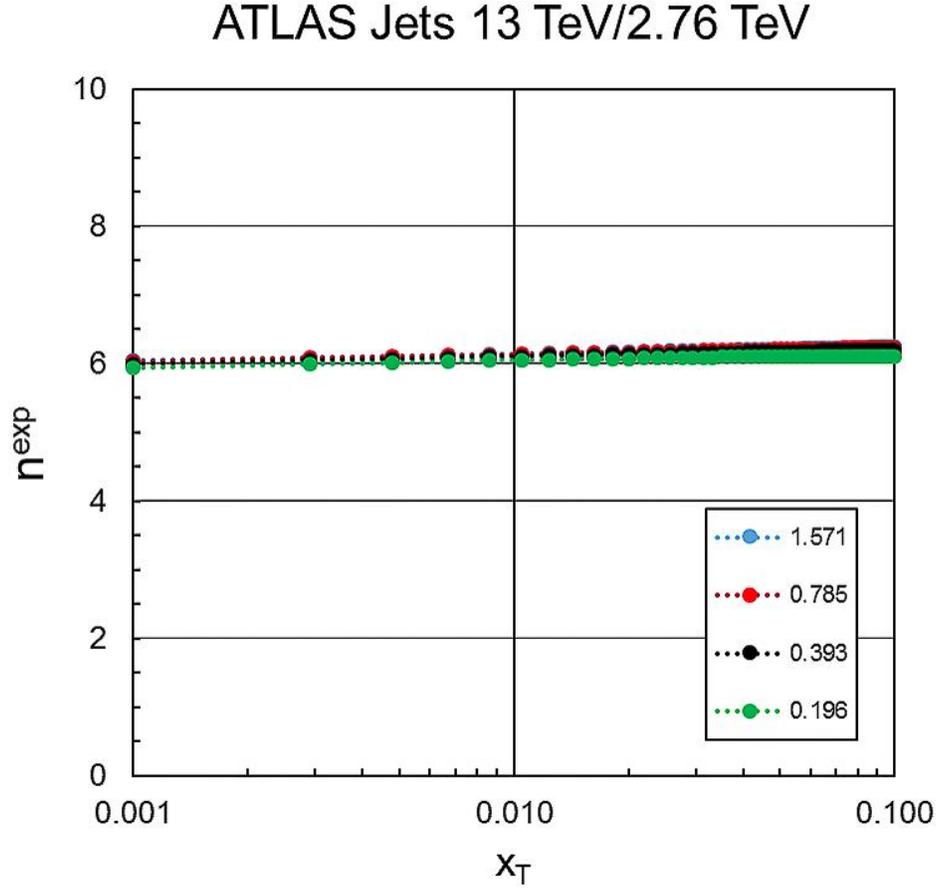

Fig. A3: The effective $p_T$ exponent analyzed by the ratio of ATLAS inclusive jets measured at 13 TeV and 2.76 TeV is plotted as a function of $x_T$ from Eq. (A5), which includes all the s-dependent terms. We see that the intrinsic $p_T$ dependence is correctly calculated.

In order to compute the true $p_T$ power exponent, n, of the invariant cross section given in Eq. (A1), we must include not only the θ dependence, or equivalently the $x_R$ dependence of the cross section, but also the α-term s-dependence. Thus, Eq. (A1) becomes:

$$\sigma^{inv} \equiv E\frac{d^3\sigma}{dp^3}(AB \to CX) = \frac{\alpha(\sqrt{s})(1-x_R)^{nx_R(\sqrt{s},pT)}}{p_T^n} \qquad (A4)$$

In the limit of small $x_T$ ($x_T = x_R \sin(\theta)$) Eq. A4 implies:

$$n^{\exp} = \frac{-\ln\left(\sigma^{inv}(x_T,\sqrt{s_1})/\sigma^{inv}(x_T,\sqrt{s_2})\right) + \ln\left(\alpha(\sqrt{s_1})/\alpha(\sqrt{s_2})\right)}{\ln\left(\sqrt{s_1}/\sqrt{s_2}\right)} \quad (A5)$$

The resultant $n^{\exp}$ is shown in Fig. A3, where it is clear that the $n^{\exp} = 6$ is regained with the necessary $\alpha(s)$-term of Eq. (A5) operative (see Fig. 11). Note:

$$\frac{\ln\left(\alpha(\sqrt{s_1})/\alpha(\sqrt{s_2})\right)}{\ln\left(\sqrt{s_1}/\sqrt{s_2}\right)} \approx 1.9 \quad (A6)$$

thereby accounting for the change $n^{\exp} \sim 4$ to $n^{\exp} \sim 6$. Note that the value of Eq. A6 is a reflection of the s-dependence of $\alpha(\sqrt{s}) \sim (\sqrt{s})^2 = s$.

Therefore, the method of Arleo, *et al.* [39] determines the effective $p_T$ power exponent of ATLAS inclusive jets to be $n_{pT} \sim 4$ (Fig. A2) because the overall s-dependence of the $\alpha$-term of the invariant inclusive cross section has been neglected (Fig. 11 and Table Ib). It is not unreasonable to conclude that the varying $p_T$ power exponents determined in Arleo, *et al.* analysis [39] result from the neglect of the s-dependences of the corresponding $\alpha$-terms. Our analysis, which determines the $p_T$ dependence at a fixed $\sqrt{s}$ by extrapolating the $(1-x_R)$ function to $x_R = 0$, finds $n_{pT} \approx 6.4 \pm 0.5$ for many inclusive measurements over a wide energy range (Tables I, II and V) and excludes $n_{pT} = 4$ by 3.8 $\sigma$.